\documentclass[aps,pra,showpacs,twoside,10pt,floatfix,nofootinbib,longbibliography,twocolumn]{revtex4-2}
\usepackage{babel} 
\usepackage[T1]{fontenc}
\usepackage[utf8]{inputenc}
\usepackage{geometry}
\usepackage{amssymb,amsfonts,amsmath,amsthm,mathtools,braket}
\usepackage{lmodern}
\usepackage{newtxtext}
\usepackage{newtxmath}
\usepackage{xcolor,comment,wrapfig}
\usepackage[normalem]{ulem}
\usepackage[colorlinks=true, citecolor=red, urlcolor=blue]{hyperref}

\newcommand{\obj}[1]{\boldsymbol{#1}}

\definecolor{boldcolor}{HTML}{703AA6}

\newcommand{\prop}[1]{\noindent\textbf{\textcolor{boldcolor}{$\blacksquare$ Proposition #1.}}}

\begin{document}

\title{Optimal performance of driven-dissipative many-body quantum batteries beyond complete nonreciprocity}
\author{Vimal Sreekanth, Chandrima B. Pushpan, Amit Kumar Pal}
\affiliation{Department of Physics, Indian Institute of Technology Palakkad, Palakkad 678 623, India}

\begin{abstract}
Proposals for nonreciprocal quantum batteries identify the point of complete
nonreciprocity, where coherent and reservoir-mediated couplings balance, as the operating point for enhanced charging. We show that, for batteries built from interacting qubits, this point is, in general, suboptimal. The optimal design is set by a more intricate interplay between the coherent coupling and the local and collective dissipation rates, of which complete nonreciprocity is only a special case. We consider a many-body battery charged by a driven many-body charger, with each qubit coupled to a local bath and the complete setup immersed in a shared engineered reservoir. For a weakly driven two-qubit setup, we prove the Dzyaloshinskii-Moriya interaction to be the optimal coupling, and show that the battery capacity is maximized at a coupling strength fixed by the local and collective dissipation rates, which coincides with complete nonreciprocity only
for a specific balance between them. The shared reservoir enhances the capacity for couplings weaker or stronger than both local dissipation rates, and can increase the extractable work even where it reduces the capacity. At strong drive, with the Dzyaloshinskii-Moriya interaction between the charger and the battery, the capacity is bounded by $1/2$. For qubits arranged in a star network with a single central battery qubit, these results remain qualitatively unchanged, while the maximum capacity of a battery charged by $N_A$ weakly driven charger qubits saturates with $N_A$ in the presence of the shared reservoir, approaching its limit as $N_A^{-1}$. In contrast, the maximum capacity of $N_B$ qubit batteries charged by a single-qubit charger decays as
$N_B^{-2}$.
\end{abstract}

\maketitle

\section{Introduction} 
\label{sec:intro}

\begin{figure*}
    \centering    \includegraphics[width=\linewidth]{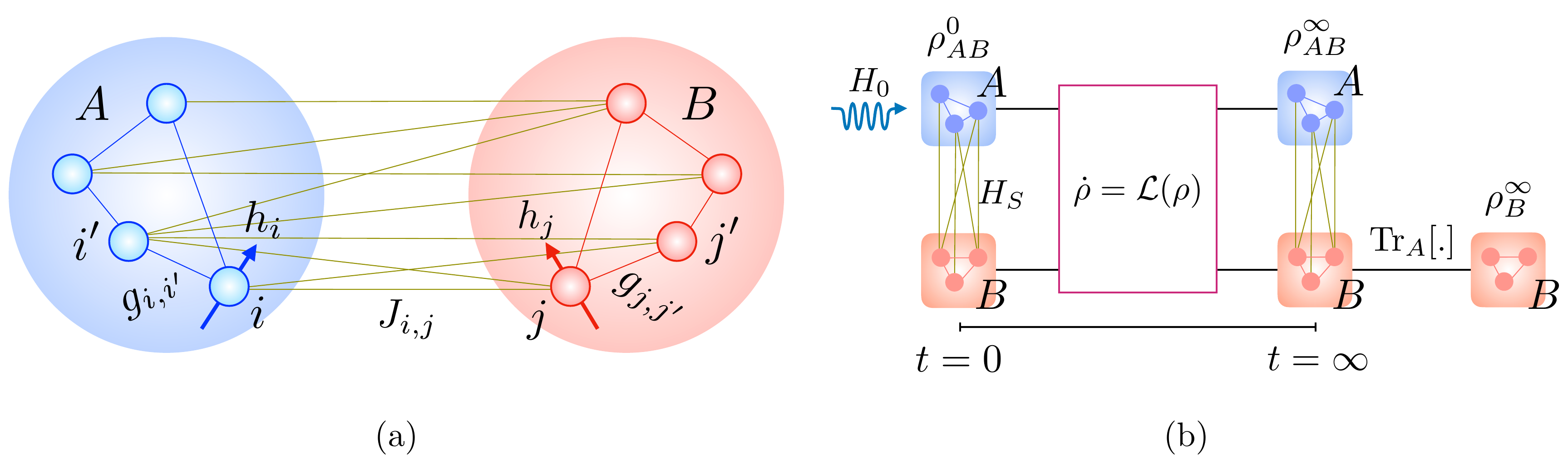}
    \caption{\textbf{(a) Interacting $N$-body system.} Schematic representation of the components $H_A$, $H_B$, and $H_{AB}$ of the system Hamiltonian $H_S$ describing  $N$ interacting DoFs belonging to either of the species $A$, or $B$. \textbf{(b) Bipartite battery-charger setup.} The setup is constituted of a many-body battery $B$ and a many-body auxiliary charger $A$. The battery $B$ is charged via an evolution of the battery-charger setup governed by the quantum master equation $\dot{\rho}=\mathcal{L}(\rho)$, followed by decoupling the charger.}
    \label{fig:setup}
\end{figure*}

In the past decades,  quantum batteries~\cite{Alicki2013,Campaioli2018,Bhattacharjee2021,Campaioli2024}, have emerged as an active area of research, with numerous proposals investigating diverse architectures and strategies for energy storage~\cite{Hovhannisyan2013,Binder2015,Campaioli2017,Santos2019,Crescente2020,Julia-Farre2020,Crescente2020,Crescente2022,Gyhm2022,Zhu2023,Downing2023,Ge2023,Gyhm2024,Song2024,Lai2024,Evangelakos2025,Andolina2025}. Important use cases such as powering a quantum computer~\cite{Kurman2026} have been introduced, along with experimental realizations of quantum batteries in trapped ions~\cite{Zhang2025}, nuclear magnetic resonance systems~\cite{Joshi2022}, quantum dots~\cite{Wenniger2023}, transmons~\cite{Hu2022,Dou2023}, nitrogen vacancy centers~\cite{Niu2024}, and organic semiconductors~\cite{Quach2022}. Within the general framework of quantum energy storage devices, bipartite  quantum batteries~\cite{Yan2023,Zhang2024a,Pushpan2026} constituted of the battery coupled to an auxiliary charger,  have been under focus, where both the battery and the charger can individually be further constituted of multiple parties. The charging protocol is typically constituted of a time evolution of the battery-charger system, initialized in a state $\rho_0$ such that the battery state $\rho_{B,0}=\text{Tr}_A\rho_0$ is a \emph{passive}  state~\cite{Pusz1978,Allahverdyan2004,Singh2021,Sen2021,Shi2022}, via a unitary evolution $\rho_t=U\rho_0 U^\dagger$, leading to an \emph{active} state $\rho_{B,t}=\text{Tr}_A\rho_{t}$ of the battery. Work extraction from $\rho_{B,t}$ can be accomplished either by performing a projective measurement on the auxiliary charger~\cite{Yan2023,Zhang2024a}, or by simply discarding it altogether~\cite{Pushpan2026,song2026}. 

Along with charging cycles designed with unitary operations~\cite{Yan2023,Zhang2024a,Pushpan2026}, open quantum batteries~\cite{Farina2019,Barra2019,Bai2020,Kamin2020,Quach2020,Gherardini2020,Tabesh2020,Carrega2020,Zhao2021,Yao2021,Mitchison2021,Yao2022,Xu2022,Tirone2023,Catalano2024,Dias2024,song2026} have also drawn attention, where the battery-charger system is in contact with its environment, and subsequently the time evolution is governed by an appropriate quantum master equation (QME)~\cite{Petruccione2007,rivas2012,Lidar2019} (cf.~\cite{Gorini1976,Lindblad1976}). 
Within the bipartite architecture of the battery-charger setup, noise has been shown to enhance charging speed and work extraction~\cite{Xu2021,Ghosh2021,Shastri2025,Purkait2026}. Further, in charging central-spin batteries with peripheral spin-chargers in a central spin model (see~\cite{Liu2021,Yang2025} for the closed counterparts of a central spin battery), the presence of local and global dissipative baths has been shown to be beneficial compared to the situation where the bath(s) is (are) absent~\cite{Liu2024}.

Recently,  based on the quantum master equation proposed in~\cite{Metelmann2015}, a design of bipartite nonreciprocal battery has been introduced~\cite{Ahmadi2024}, where unidirectional energy transfer from the charger to the battery is realized via a balance, $|J|=\Gamma/2$, between the battery-charger coherent interaction of strength $J$, and the collective dissipation rate, $\Gamma$.  At this \emph{nonreciprocal} point, a fourfold advantage in the presence of the shared engineered reservoir is pointed out in comparison with the situation where the shared bath is absent. This enterprise has been extended by  engineering enhanced charging~\cite{Ahmadi2025,Khan2026}, by introducing additional components such as a quantum mediator~\cite{Sun2025} or a lossy auxiliary
cavity~\cite{Zafar2026} between the battery and the charger, and by using different alternatives for the  coherent drive~\cite{Xu2025arX,Yang2026arX}. The model has also been scaled up to multimode
cells~\cite{Khan2025}, cascaded multibattery
chains~\cite{Zhao2025}, and waveguide-array
networks~\cite{Guo2025}, extending the investigation into cascaded versus parallel network
topologies~\cite{Liu2026arX} and proposals for reservoir-engineered non-Hermitian schemes to improve the robustness of nonreciprocal charging~\cite{Lin2026}. A regime-independent sixteenfold advantage due to the shared reservoir has also been recently proposed~\cite{Khan2026arX} via  a modification of the quantum master equation proposed in~\cite{Metelmann2015}, leaving the quantification of the nonreciprocal benefit as an open question in the literature.

The advantage of nonreciprocal charging has so far been quantified almost exclusively \emph{at} the nonreciprocal point, where an enhancement of the stored energy over conventional charger--battery systems without a shared
bath has been reported~\cite{Ahmadi2024,Ahmadi2025,Khan2026,Khan2026arX}. It remains unclear, however, whether complete nonreciprocity is actually \emph{required}
to reach the \emph{optimal} enhancement, or whether a partially reciprocal coupling can perform as well, or even better. Answering this calls for optimizing the battery performance over the full reciprocity landscape, together with a systematic comparison against the case in which the collective reservoir is absent. In particular, the explicit conditions on the local and collective dissipation rates, the driving strength, and the coherent coupling that optimize
the performance of a driven-dissipative battery have not yet been identified.

Further, existing proposals for the design and performance of nonreciprocal quantum
batteries are formulated predominantly in terms of linearly coupled harmonic
oscillators~\cite{Metelmann2015,Ahmadi2024,Ahmadi2025,Khan2026,Sun2025}. In
such Gaussian settings, the equations of motion for the first and second
moments of the mode operators close exactly. As a result, the battery-energy dynamics at the nonreciprocal point $|J|=\Gamma/2$, as well as the full steady state, follow from a small set of coupled linear equations. However, there is no guarantee that these advantages survive when the charger and the battery are built from interacting finite-dimensional degrees of freedom, such as qubits arranged on a lattice, where the moment hierarchy no longer closes and interactions can compete with, or even undermine, directional energy flow. This question is timely, since directional interactions can now be engineered in
experimental platforms such as superconducting circuits~\cite{Ren2025}, and
nonreciprocal quantum many-body systems are attracting rapidly growing
attention~\cite{Kirton2019,Chiacchio2023,Zhang2026}, with a recent realization of nonreciprocal battery in trapped-ion systems~\cite{Zhang2026experimental}.

Motivated by these questions, in this paper, we determine the conditions for the optimal performance of a driven-dissipative many-body quantum battery over the full reciprocity landscape. We work within the bipartite battery-charger setup, where both the charger and the battery are constituted of multiple degrees of freedom (DoFs), each connected to its own local dissipative environment, while the whole setup is immersed in a shared dissipative reservoir. We focus on \emph{charger-mediated}, or indirect, charging of the battery~\cite{Andolina2018,Farina2019}, where none of the DoFs of the battery are externally driven. We quantify the capacity of the battery in terms of its energy, ergotropy~\cite{Pusz1978,Allahverdyan2004,Singh2021,Sen2021,Shi2022}, and daemonic ergotropy~\cite{Francica2017,Bernards2019,Morrone2023,Barra2024,Kua_2026}, and introduce the \emph{bound capacity}, corresponding to the stored energy that cannot be extracted as work. Considering the DoFs to be qubits arranged on an arbitrary lattice, we prove that the steady state of the battery-charger setup is unique, so that the steady-state performance of the battery is independent of its initial state. For a single-qubit battery charged by a single-qubit charger, we prove that, in the limit where the drive strength is small compared to the coherent battery-charger coupling, the steady-state capacity of the battery, as well as its advantage over the collective-reservoir-free (CRF) limit, is maximized when the battery and the charger interact via the Dzyaloshinskii-Moriya (DM) interaction~\cite{Anderson1959,Dzyaloshinsky1958,Moriya1960,Moriya1960a} (see also~\cite{Lewenstein2007,Amico2008}). Developing a perturbative approach in the drive strength, we show that the optimal coupling strength is determined by the local and collective dissipation rates, and coincides with the nonreciprocal point only for a specific balance between the local and collective dissipation rates, implying that \emph{complete nonreciprocity is, in general, not required for optimal charging}. We further show that the maximum capacity is bounded above by $1/4$, and derive explicit conditions under which the shared reservoir enhances the capacity: an advantage is obtained when the coupling is either weaker or stronger than both local dissipation rates, while an intermediate coupling, which contains the capacity optimum, is detrimental. In the opposite limit of a strong drive, we show that the capacity of the battery is upper bounded by $1/2$, which is reached only for an asymptotically strong coupling, while an advantage due to the shared reservoir is obtained only for a weak coupling, so that the optimal capacity and the advantage due to the shared reservoir are not achieved simultaneously.

We extend our analysis to a star network of qubits, where $N_A\,(>1)$ identically and weakly driven charger qubits charge a single central battery qubit. We prove the DM interaction between the battery and a charger qubit to remain the optimal choice, and obtain explicit conditions on the local and collective dissipation rates and the coupling strength for an advantage due to the shared reservoir. We show that, in the presence of the shared reservoir, the maximum capacity of the battery grows with $N_A$ and saturates, as $N_A \to \infty$, to a value determined by the drive and the bath parameters, approaching it as $\sim N_A^{-1}$. In contrast, in the absence of the shared reservoir, the maximum capacity grows linearly with $N_A$. In the complementary setup, where a single central charger qubit charges $N_B$ battery qubits, the maximum capacity decays as $\sim N_B^{-2}$.

The rest of the paper is organized as follows. In Sec.~\ref{sec:model_and_methodology}, we formally introduce the bipartite quantum many-body system in the presence of local as well as collective dissipation and an external drive on one subsystem. We also discuss the suitable frame in which the rest of the calculations are performed, and describe the figures of merit for the battery performance. In Sec.~\ref{sec:charging_a_qubit}, we introduce the minimal battery-charger setup where both the battery and the charger are constituted of one qubit each, and discuss the performance of the battery. We develop a perturbative approach to probe the case of the weakly-driven charger, and extend the result beyond the perturbative regime within specific parameter ranges. We extend the calculations to a setup of multiple charger qubits with a single battery qubit, and a single charger qubit with multiple battery qubits in Sec.~\ref{sec:multiqubit_charger}, and discuss the corresponding results and scaling, along with extension of the results beyond the perturbative regime.  Sec.~\ref{sec:conclusion} contains the concluding remarks and outlook.

\section{Setting up the many-body quantum battery}
\label{sec:model_and_methodology}

We now formally introduce the driven-dissipative many-body system. We also introduce the setup of the battery within the system, and discuss the parameters used to quantify the battery performance.   

\subsection{A driven-dissipative many-body system}
\label{subsec:system}

Let us consider a system of $N$ interacting degrees of freedom (DoFs) belonging to either of two species, namely, $A$, and $B$, situated on the vertices of a bipartite graph $G$. For specificity, let us assume that the species $A$ ($B$) contains $N_A$ ($N_B$) DoFs, denoted by the index $i=1,2,\cdots,N_A$ ($j=1,2,\cdots,N_B$) and each having a local Hilbert space dimension $d_A$ ($d_B$) with $N_A+N_B=N$. An intra-species link between $i$  and $i^\prime$  ($j$ and $j^\prime$), denoted by $\langle i,i^\prime\rangle$ ($\langle j,j^\prime\rangle$) with $i\neq i^\prime$ ($j\neq j^\prime$) exists in $A$ ($B$) if the corresponding DoFs interact, while the inter-species link $\langle i,j\rangle$ between the vertices $i\in A$ and $j\in B$ exists in $G$ if the DoF on $i$ interacts with the DoF on $j$ (see Fig.~\ref{fig:setup}(a)).   

The Hamiltonian representing the interaction between these DoFs is given by  
\begin{eqnarray}
    H_S=H_0+H,
\end{eqnarray}
with $H_0$ being a driving term corresponding to subsystem $A$ only, and $H=H_{A}+H_{B}+H_{AB}$, where the components $H_A$ and $H_B$ describe the intrinsic dynamics within species, and $H_{AB}$ is responsible for mediating coherent exchange leading to transport between $A$ and $B$. Specifically, we consider
\begin{eqnarray}
\label{eq:system_hamiltonian}
    H_A &=&\sum_{i\in A}h_{i}a_i^+a_i^-+ \sum_{\langle i,i^\prime\rangle\in G}\left(g_{i,i^\prime}a_i^+a_{i^\prime}^- +g_{i,i^\prime}^*a_i^-a_{i^\prime}^+\right),\nonumber\\
    H_B &=&\sum_{j\in B}h_{j}b_j^+b_j^-+ \sum_{\langle j,j^\prime\rangle\in G}\left(g_{j,j^\prime}b_j^+b_{j^\prime}^- +g_{j,j^\prime}^*b_j^-b_{j^\prime}^+\right),\nonumber\\
    H_{AB} &=& \sum_{\langle i,j\rangle\in G}\left[J_{i,j}a_i^+b_j^- +J^*_{i,j}a_i^-b_j^+\right],    
\end{eqnarray}
where  $a^-$ and $b^-$ ($a^+$ and $b^+$) are the lowering (raising) operators corresponding respectively to the species $A$  and $B$, satisfying 
\begin{eqnarray}
    a_i^{\pm}=\left(a_i^{\mp}\right)^\dagger,\; b_j^{\pm}=\left(b_j^{\mp}\right)^\dagger,\;
\end{eqnarray}
as well as
\begin{eqnarray}
    [a_i^+a_{i}^-,a_{i^\prime}^\pm]=\pm\delta_{i,i^\prime}a_{i^\prime}^\pm,\;[b_j^+b_{j}^-,b_{j^\prime}^\pm]=\pm\delta_{j,j^\prime}b_{j^\prime}^\pm,
\end{eqnarray}
while $[a_i^\pm,b_j^\pm]=0$. The parameters $h_{i}$, $h_{j}$ ($\in\mathbb{R}$) represent field strengths local to the DoFs, $g_{i,i^\prime}$ and $g_{j,j^\prime}$, given by 
\begin{eqnarray}
g_{i,i^\prime}=|g_{i,i^\prime}|\text{e}^{\text{i}\nu_{i,i^\prime}}, \;\; g_{j,j^\prime}=|g_{j,j^\prime}|\text{e}^{\text{i}\nu_{j,j^\prime}}
\end{eqnarray}
are the complex intra-species pairwise hopping strengths in $A$ and in $B$, respectively (see Fig.~\ref{fig:setup}), and
\begin{eqnarray}\label{eq:J}
    J_{i,j}=|J_{i,j}|\text{e}^{\text{i}\theta_{i,j}},
\end{eqnarray} 
is the complex inter-species hopping strength~\cite{Jimenez-Garcia2012,Carusotto2013,Lienhard2020}.  We further consider $H_0$ to be an external coherent drive on $A$, given by 
\begin{eqnarray}\label{eq:driving}
    H_0 &=& \sum_{i\in A}F_ia_i^++F_i^*a_i^-,
\end{eqnarray}
where $F_i=|F_i|\text{e}^{\text{i}\omega_i t}$ is the complex drive amplitude on the $i$th DoF in $A$, with  the frequency $\omega_i$ determining the direction of the drive on the $xy$-plane.

We assume  that each DoF in the system is subjected to local as well as collective dissipation arising from coupling to external environments. The dynamics of the system is described by the Gorini-Kossakowski-Lindblad-Sudarshan (GKLS) quantum master equation (QME)~\cite{Petruccione2007,rivas2012,Lidar2019} (cf.~\cite{Gorini1976,Lindblad1976}) (assuming $\hbar=1$)
\begin{eqnarray}\label{eq:QME}
    \frac{d\rho}{dt}&=& \mathcal{L}(\rho),
\end{eqnarray}
with 
\begin{eqnarray}
    \mathcal{L}(\rho)&=&-\text{i}[H_S,\rho]+\mathcal{D}(\rho),
    \label{eq:cal_L}
\end{eqnarray}
and 
\begin{eqnarray}
    \mathcal{D}(\rho)=\sum_{i\in A}\gamma_i\mathcal{D}_i(\rho)+\sum_{j\in B}\gamma_{j}\mathcal{D}_j(\rho)+\Gamma \mathcal{D}_{AB}(\rho),
    \label{eq:cal_D}
\end{eqnarray}
where $\gamma_i,\gamma_j$, and $\Gamma$ ($\in \mathbb{R}$) are respectively the local and collective dissipation rates, and $\mathcal{D}_\alpha(.)$ are given by  
\begin{eqnarray}
    \mathcal{D}_\alpha(\rho)=L_\alpha \rho L_\alpha^\dagger-\frac{1}{2}\{L_\alpha^\dagger L_\alpha,\rho\},\;\alpha=i,j,AB,
\end{eqnarray} 
with $L_\alpha$ being the local ($\alpha=i,j$) and global $(\alpha=AB)$ jump operators. In the limit 
\begin{eqnarray}
    \max\{|J_{i,j}|,|g_{i,i^\prime}|,|g_{j,j^\prime}|,|F_i|\}\ll\min\{h_i,h_j\},\label{eq:weakly_coupled_dof}
\end{eqnarray}
the jump  operators can be taken to be of the following form:
\begin{eqnarray}\label{eq:global_jump_operator}
L_i&=&a_i^-,\nonumber\\
L_j&=&b_j^-,\nonumber\\
L_{AB}&=&\sum_{i\in A}\mathcal{A}_ia_i^-+\sum_{j\in B}\mathcal{B}_jb_j^-,
\end{eqnarray}
where $\mathcal{A}_i$ and $\mathcal{B}_j$ are complex in general, encompassing a variety of examples~\cite{Carmichael1993,Metelmann2015,Pichler2015,Masson2022,Forero2025,Ren2025,Zhang2026experimental}.

\subsection{Choosing the frame}
\label{subsec:frame}

We  choose to work in the rotated frame obtained due to the unitary operation
\begin{eqnarray}
    U=\exp\left[{\text{i}\left(\sum_{i\in A}h_ia_i^+a_i^-+\sum_{j\in B}h_jb_j^+b_j^-\right)t}\right],
\end{eqnarray}
with $\tilde{\rho}=U\rho U^\dagger$, such that
\begin{eqnarray}
    \frac{d\tilde{\rho}}{dt}
    &=&-\text{i}[\tilde{H}_S,\tilde{\rho}]+\tilde{\mathcal{D}}(\rho),
    \label{eq:rotated_qme_initial}
\end{eqnarray}
with 
\begin{eqnarray}
    \tilde{H}_S&=&UH_S U^\dagger=\tilde{H}_0+\tilde{H}_A+\tilde{H}_B+\tilde{H}_{AB},
\end{eqnarray}
where $\tilde{\mathcal{D}}(\rho)=U\mathcal{D}(\rho)U^\dagger$. Explicit calculations lead to  
\begin{eqnarray}
    \tilde{H}_0&=&\sum_{i\in A}|F_i|\left[\text{e}^{\text{i}(\omega_i+h_i)t}a_i^++\text{e}^{-\text{i}(\omega_i+h_i)t}a_i^-\right],\nonumber\\
    \tilde{H}_A&=&\sum_{\langle i,i^\prime\rangle\in G}\big[g_{i,i^\prime}\text{e}^{\text{i}h_{i,i^\prime} t}a_i^+ a_{i^\prime}^- +g_{i,i^\prime}^*\text{e}^{-\text{i}h_{i,i^\prime}t}a_i^- a_{i^\prime}^+\big],\nonumber\\
    \tilde{H}_B&=&\sum_{\langle j,j^\prime\rangle\in G}\big[g_{j,j^\prime}\text{e}^{\text{i}h_{j,j^\prime}t}b_j^+ b_{j^\prime}^- +g_{j,j^\prime}^*\text{e}^{-\text{i}h_{j,j^\prime}t}b_j^- b_{j^\prime}^+\big],\nonumber\\
    \tilde{H}_{AB}&=&\sum_{\langle i,j\rangle\in G} \big[J_{i,j}e^{\text{i}h_{i,j}t}a_i^+b_j^-+J_{i,j}^*e^{-\text{i}h_{i,j}t}a_i^-b_j^+\big],\nonumber\\
\end{eqnarray} 
where $h_{i,j}=h_i-h_{j}$, and 
\begin{eqnarray}
    \tilde{L}_i&=&\text{e}^{-\text{i}h_it}L_i, \nonumber\\
    \tilde{L}_j&=&\text{e}^{-\text{i}h_jt}L_j,\nonumber\\
    \tilde{L}_{AB}&=&\sum_{i\in A}\mathcal{A}_i\text{e}^{-\text{i}h_it}a_i^-+\sum_{j\in B}\mathcal{B}_j\text{e}^{-\text{i}h_jt}b_j^-, 
\end{eqnarray}
where $\tilde{L}_{\alpha}=UL_\alpha U^\dagger$, $\alpha=i,j,AB$. Choosing $h_i=h_j=h$ $\forall i\in A,j\in B$, and $\omega_i=-h$ $\forall i\in A$, which we maintain throughout this paper, one further obtains 
\begin{eqnarray}
    \tilde{H}_A&=&\sum_{\langle i,i^\prime\rangle\in G}\big[g_{i,i^\prime}a_i^+ a_{i^\prime}^- +g_{i,i^\prime}^*a_i^- a_{i^\prime}^+\big],\nonumber\\
    \tilde{H}_B&=&\sum_{\langle j,j^\prime\rangle\in G}\big[g_{j,j^\prime}b_j^+ b_{j^\prime}^- +g_{j,j^\prime}^*b_j^- b_{j^\prime}^+\big],
\end{eqnarray}
and 
$\tilde{H}_{AB}=H_{AB}$,  with 
\begin{eqnarray}
    \tilde{H}_0= \sum_{i\in A}|F_i|(a_i^++a_i^-),
\end{eqnarray}
and $\tilde{\mathcal{D}}(\rho)=\mathcal{D}(\tilde{\rho})$. This leads to 
\begin{eqnarray}
    \frac{d\tilde{\rho}}{dt}
    &=&-\text{i}[\tilde{H}_S,\tilde{\rho}]+\mathcal{D}(\tilde{\rho})=\tilde{\mathcal{L}}(\tilde{\rho}),
    \label{eq:rotated_qme_final}
\end{eqnarray}
and further reduces Eq.~(\ref{eq:weakly_coupled_dof}) to 
\begin{eqnarray}
    \max\{|J_{i,j}|,|g_{i,i^\prime}|,|g_{j,j^\prime}|,|F_i|\}\ll h.\label{eq:final_approx}
\end{eqnarray} 
The expectation value $\left\langle\mathcal{O}\right\rangle$ of an arbitrary time-independent operator $\mathcal{O}$ can be derived as a function of $t$ from Eq.~(\ref{eq:rotated_qme_final}) as 
\begin{eqnarray}
\label{eq:Opeator_evolution_gen_1}
    \frac{d\left\langle\mathcal{O}\right\rangle}{dt}&=&
    -\text{i}\text{Tr}\big[[\tilde{H}_0+\tilde{H}_A+\tilde{H}_B+H_{AB},\tilde{\rho}]\mathcal{O}\big]\nonumber\\&&+\sum_{i\in A}\gamma_i \text{Tr}\left[\mathcal{D}_i(\tilde{\rho})\mathcal{O}\right]\nonumber\\&&
    +\sum_{j\in B}\gamma_j\text{Tr}\left[\mathcal{D}_j(\tilde{\rho})\mathcal{O}\right]+\Gamma\text{Tr}\left[\mathcal{D}_{AB}(\tilde{\rho})\mathcal{O}\right],\nonumber\\
    &=&\text{Tr}\left[\tilde{\rho}\tilde{\mathcal{L}}^\dagger(\mathcal{O})\right]=\langle\tilde{\mathcal{L}}^\dagger(\mathcal{O})\rangle.
\end{eqnarray}

\subsection{The battery and its performance parameters}
\label{subsec:performance}

We consider a many-body quantum battery (QB) in the form of a subsystem of the driven-dissipative many-body system. In this paper, we are specifically interested in \emph{charger-mediated}, or \emph{indirect} charging of the QB~\cite{Andolina2018, Farina2019}, where none of the DoFs constituting the QB is externally driven. This motivates the choice of the subsystem $B$ as the battery, which is charged indirectly via the auxiliary charger subsystem $A$ through the charging dynamics described by the QME (Eq.~(\ref{eq:QME})). The charger is decoupled from the battery at a time $t$ (see Fig.~\ref{fig:setup}(b)), when the state of the battery-charger system, $\rho(t)$, is obtained as the solution of Eq.~(\ref{eq:QME}).

Without loss of generality, we assume the ground state energy of the battery to be zero, while $E^{\max}$ is the energy of the battery corresponding to the highest energy eigenstate, and $E=\text{Tr}\left[H_B\tilde{\rho}_B(t)\right]$ is the instantaneous average  energy of the battery, with $E^{\infty}=E(t=\infty)$ denoting the average stored energy in the steady state. In the presence of the collective reservoir, we define the \emph{instantaneous charging capacity} (ICC) of the battery at time $t$ as 
\begin{eqnarray}
    C(t)=\frac{E(t)}{E^{\max}},
    \label{eq:cc}
\end{eqnarray}
while the same in the \emph{collective-reservoir-free} (CRF) limit is denoted by $C_{0}(t)$, the ``$0$" in the subscript denoting $\Gamma=0$. We specifically focus on the effect of the presence of the collective reservoir in the battery capacity, quantified by 
\begin{eqnarray}
    \delta C(t)=C(t)-C_{0}(t),
\end{eqnarray}
with $\left|\delta C(t)\right|\leq 1$. 
A similar figure of merit can also be proposed in terms of the \emph{ergotropy}, i.e., maximum energy extractable from the battery~\cite{Pusz1978,Allahverdyan2004,Singh2021,Sen2021,Shi2022} (see Appendix~\ref{app:ergotropy_definitions} for definitions), denoted by $\mathcal{E}$, where one can define 
\begin{eqnarray}
    W(t)&=&\frac{\mathcal{E}(t)}{\mathcal{E}^{\max}},\;  W_0(t)=\frac{\mathcal{E}_0(t)}{\mathcal{E}^{\max}_0},
\end{eqnarray}
and 
\begin{eqnarray}
    \delta W(t)=W(t)-W_{0}(t),
\end{eqnarray}
where one can identify $\mathcal{E}^{\max}=E^{\max}$. Further, $\overline{W}(t)$ and $\delta\overline{W}(t)$ can also be defined using \emph{daemonic ergotropy}~\cite{Francica2017,Bernards2019,Morrone2023,Barra2024,Kua_2026} (see Appendix~\ref{app:ergotropy_definitions} for definitions), $\overline{\mathcal{E}}$, in a similar fashion, where $\overline{\mathcal{E}}^{\max}=\mathcal{E}^{\max}=E^{\max}$. 

We now define the \emph{bound capacity} quantifying the fraction of energy that \emph{cannot} be extracted from the battery as  
\begin{eqnarray}
    C^{\mathrm{b}}(t)=C(t)-W(t)\geq 0,
\end{eqnarray}
which, in the CRF limit, is given by $C^{\mathrm{b}}_0(t)$. We further define    
\begin{eqnarray}
    \delta C^{\mathrm{b}}(t)=C^{\mathrm{b}}(t)-C^{\mathrm{b}}_0(t),
\end{eqnarray}
and note that 
\begin{eqnarray}
    \delta C^{\mathrm{b}}(t)\geq 0\implies\delta C(t)\geq \delta W(t).
\end{eqnarray}
Since $\overline{W}(t)\geq W(t)$, one can always reduce the bound capacity by choosing daemonic ergotropy over ergotropy to extract work from the battery. 

In this paper, we specifically focus on the situation where the individual DoFs correspond to spin-$1/2$ particles (qubits), with 
$a^{\pm}, b^{\pm}$ identified as spin raising and lowering operators, and 
\begin{eqnarray}
    a^{z}&=&[a^+,a^-], a^{x}=a^++a^-, a^y=-\text{i}(a^+-a^-),\nonumber\\
    b^{z}&=&[b^+,b^-], b^{x}=b^++b^-, b^y=-\text{i}(b^+-b^-),
    \label{eq:pauli}
\end{eqnarray}
are the standard Pauli matrices corresponding to the two species of spins. Further, we investigate the performance of the battery in the steady state of the battery-charger setup, given by $\tilde{\rho}_\infty=\tilde{\rho}(t=\infty)$, where the performance is quantified using  $\tilde{\rho}_B^\infty=\tilde{\rho}_B(t=\infty)$, with $\tilde{\rho}_B(t)=\text{Tr}_A\left[\tilde{\rho}(t)\right]$, and subsequently the relevant capacities can be computed.

Note that $\tilde{\rho}_\infty$ may, in principle, depend on the choice of the initial state $\tilde{\rho}(0)$ of the battery-charger setup, and subsequently on the initial state $\tilde{\rho}_B(0)$ of the battery. However, for a system with a finite-dimensional Hilbert space, each super-operator $\mathcal{L}(.)$ corresponds to at least one stationary state~\cite{Baumgartner2008a,Baumgartner2008b}, i.e., $\tilde{\rho}(0)=\tilde{\rho}_\infty$. Further, the steady states, $\tilde{\rho}_\infty$, of
\emph{irreducible} systems are unique and faithful~\cite{Zhang2024}. We consider the battery-charger setups constituted of qubits on arbitrary graphs, for which the irreducibility can be explicitly proved for 
\begin{eqnarray}
    \gamma_i,\gamma_j&\neq&0,\forall i\in A,j\in B\nonumber\\
    \text{i}J^*_{i,j}+\frac{\Gamma}{2}&\neq& 0, \forall \langle i,j\rangle\in G,
    \label{eq:irreducibility_validity}
\end{eqnarray}
such that $\tilde{\rho}_\infty$ is unique, thereby avoiding the initial state dependence (see Appendix~\ref{app:steady_state_existence} for the proof of irreducibility in cases of $|F_i|\neq 0$ $\forall i\in A$, and Appendix~\ref{app:perturbation} for the case of $|F_i|=0$ $\forall i\in A$). Further, for simplicity, we assume $|F_i|=|F|$ and $\gamma_i=\gamma_A$ $\forall i\in A$, $\gamma_j=\gamma_B$ $\forall j\in B$, $J_{i,j}=J$ $\forall \langle i,j\rangle\in G$, $g_{i,i^\prime}=g_A$ $\forall \langle i,i^\prime\rangle\in G$, and $g_{j,j^\prime}=g_B$ $\forall \langle j,j^\prime\rangle\in G$. We further assume $0\leq \gamma_A,\gamma_B,\Gamma\leq 1$ unless otherwise mentioned.

\begin{figure}
    \centering
    \includegraphics[width=\linewidth]{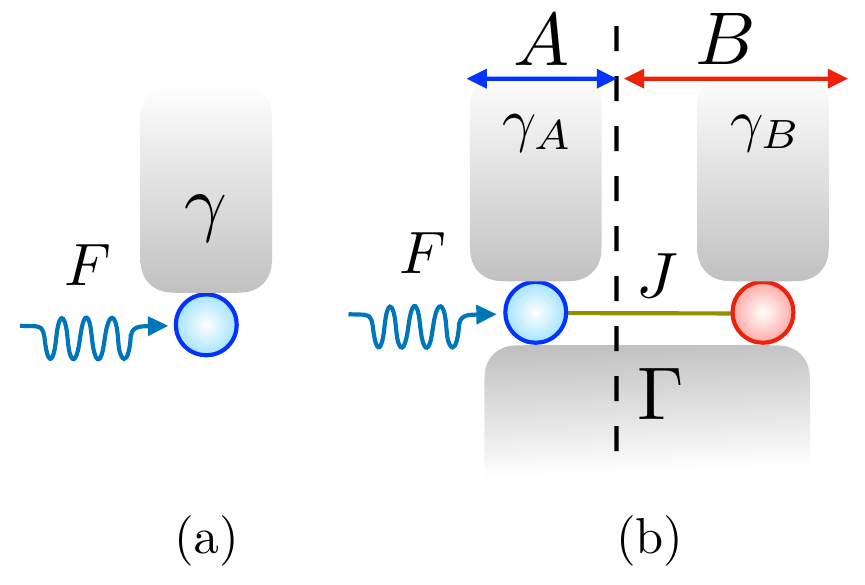}
    \caption{\textbf{Direct and indirect charging of a qubit.} (a) A qubit, in contact with a dissipative bath with a dissipation rate $\gamma$ is directly charged by an external drive of strength $F$. (b) A qubit, labeled by $B$, is indirectly charged via a second qubit, labeled by $A$ and playing the role of a charger, which is subjected to the external drive $F$. The qubits are connected by a coherent interaction of strength $J$, and are immersed in a common dissipative bath with a dissipation rate $\Gamma$, while being individually connected to two local dissipative reservoirs with dissipation rates $\gamma_{A}$ and $\gamma_B$.}
    \label{fig:simple_battery}
\end{figure}

\section{Two-qubit battery-charger setup}
\label{sec:charging_a_qubit}

We first consider the charging of a single qubit within the framework discussed in Sec.~\ref{sec:model_and_methodology}. Note that a simple yet realistic model of \emph{directly charging} a qubit corresponds to a single qubit in this framework, which consists of an externally driven qubit connected to a dissipative environment (see Fig.~\ref{fig:simple_battery}(a)). The existence and uniqueness of the steady state of this driven-dissipative single-qubit system is discussed in Appendix~\ref{app:irreducibility_single_qubit}. Since the system is constituted of only one qubit, we discard the labeling of the species for brevity.

Using Eq.~(\ref{eq:Opeator_evolution_gen_1}) (see Sec.~\ref{subsec:frame}), one obtains
\begin{eqnarray}
    \frac{d\langle a^z\rangle}{dt}&=&-2\text{i}|F|\left[\langle a^+\rangle-\langle a^-\rangle\right]-\gamma\left[\langle a^z\rangle+1\right],\nonumber\\
    \frac{d\langle a^+\rangle}{dt}&=&-\text{i}|F|\langle a^z\rangle-\frac{\gamma}{2}\langle a^+\rangle,\nonumber\\
    \frac{d\langle a^-\rangle}{dt}&=&\text{i}|F|\langle a^z\rangle-\frac{\gamma}{2}\langle a^-\rangle,
\end{eqnarray}
where $a^\pm$ are as defined in Eq.~(\ref{eq:pauli}).  Solving,  
\begin{eqnarray}
    E^\infty&=&h\langle a^+a^-\rangle=\frac{h}{2}\left(\langle a^z\rangle+1\right)=\frac{4h|F|^2}{8|F|^2+\gamma^2},
\end{eqnarray}
in the steady state. Identifying $E^{\max}=h$, one obtains
\begin{eqnarray}
    C^\infty=\frac{4}{8+(\gamma/|F|)^2},
    \label{eq:cc_single_qubit}
\end{eqnarray}
which straightforwardly leads to the following proposition. 

\prop{1} \emph{For a driven-dissipative qubit with a drive strength $F$ and a dissipation rate $\gamma$,}
\begin{eqnarray}
    C^\infty\left\{\begin{array}{lll}
       \approx  & 1/2 &\text{for } \gamma/|F|\ll 1,\\
       \leq  & 4/9 & \text{for } \gamma/|F|\geq 1,\\
       \approx & 0 &\text{for }\gamma/|F|\gg 1.
       \end{array}\right.
       \label{eq:cc_single_qubit_limits}
\end{eqnarray}

We now consider an indirect charging of a target qubit labeled by $B$ (i.e., $N_B=1$), where the externally driven charger is also constituted of a single qubit (i.e., $N_A=1$)  (see Fig.~\ref{fig:simple_battery}(b)).  In the CRF limit with $\gamma_A/|F|\ll 1$, the steady-state capacity of the qubit $A$ $\approx 1/2$, while the battery $B$ also achieves saturation ($C_{0}^\infty\approx 1/2$) fast via its channel to the saturated charger $A$ due to the coherent exchange interaction, leading to $\delta C^{\infty}\approx 0$.  This feature remains unchanged even in the presence of the collective reservoir ($0<\Gamma\leq 1$), as confirmed also by our numerical investigations. However, for $|\gamma_A|/|F|>1$, the charger and hence the battery in the CRF limit are unable to achieve saturation. Introduction of an engineered collective reservoir provides an additional channel between the battery and the charger, where appropriate tuning of the reservoir may result in $\delta C^\infty >0$. In the following, we demonstrate that this is indeed the case.

\begin{table*}[t]
\caption{Conditions for advantage over the CRF limit at fixed $|J|$ for $\Gamma>0$.}
\label{tab:dC_sign}
\begin{ruledtabular}
\begin{tabular}{lll}
Regime & Condition & Sign of $\delta C^{\infty}$\\
\hline
Weak coupling & $2|J|<\min(\gamma_A,\gamma_B)$ & $\delta C^{\infty}>0$\\
Intermediate coupling & $\min(\gamma_A,\gamma_B)<2|J|<\max(\gamma_A,\gamma_B)$ & $\delta C^{\infty}<0$\\
Strong coupling & $2|J|>\max(\gamma_A,\gamma_B)$ & $\delta C^{\infty}>0$\\
Boundaries & $2|J|=\gamma_A$ or $2|J|=\gamma_B$ & $\delta C^{\infty}=0$\\
Symmetric limit & $\gamma_A=\gamma_B=\gamma$ & $\delta C^{\infty}\geq0$ for all $|J|$, zero only at $2|J|=\gamma$\\
\end{tabular}
\end{ruledtabular}
\end{table*}

\begin{figure}
    \centering
    \includegraphics[width=\linewidth]{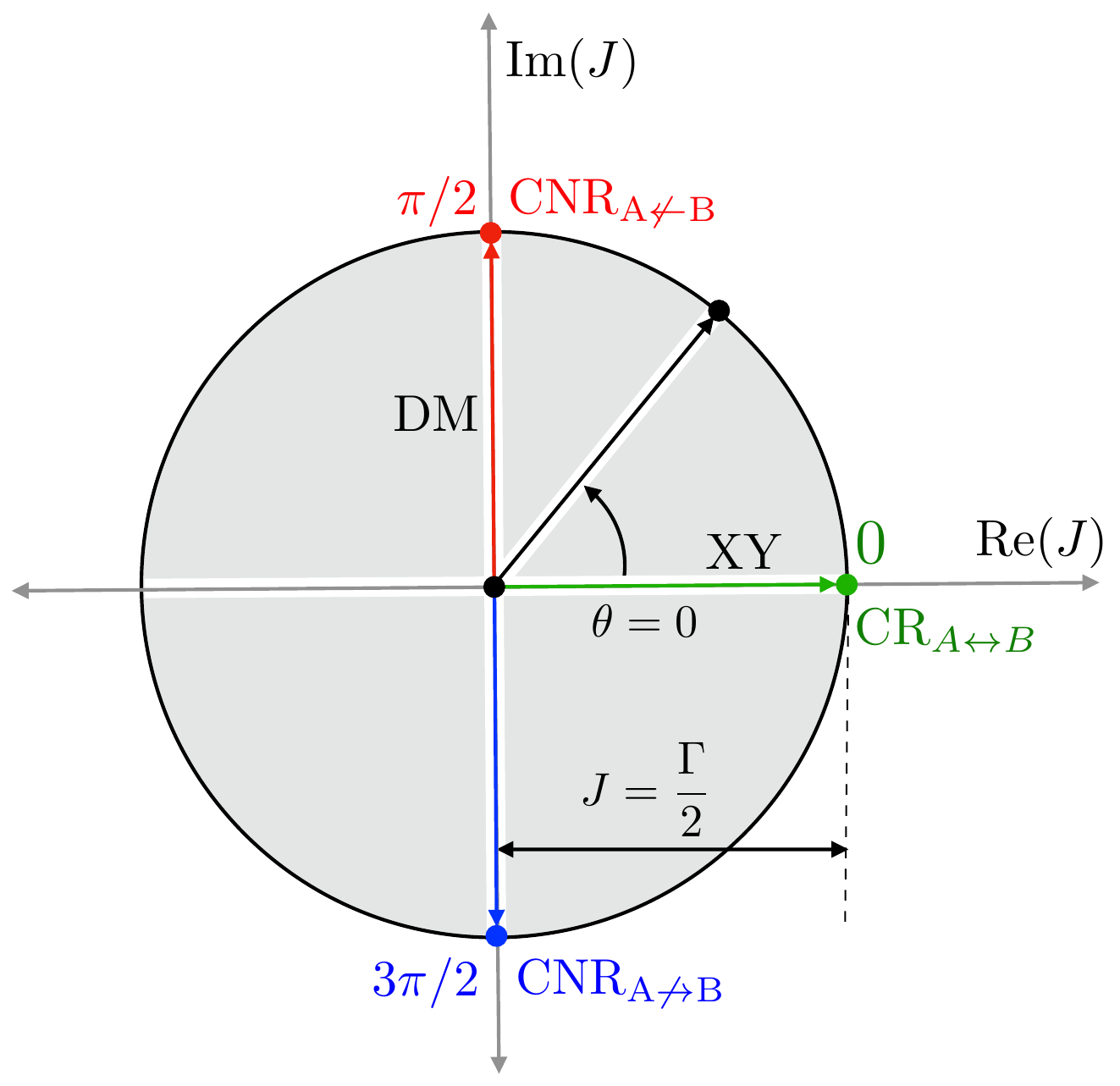}
    \caption{\textbf{Completely nonreciprocal and completely reciprocal points.} Assuming $\phi=0$,  with $|J|=\Gamma/2$, the CNR points are $\theta=\pi/2$ ($\text{CNR}_{A\not\leftarrow B}$) and  $3\pi/2$ ($\text{CNR}_{A\not\to B}$), while the $\text{CR}_{A\leftrightarrow B}$ point is given by $\theta=0$, with a choice of $n=0$. The XY model is represented by $\text{Im}(J)=0$, while $\text{Re}(J)=0$ corresponds to the DM interaction between the qubits $A$ and $B$.}\label{fig:parameter_space}
\end{figure}

To understand the effect of the battery and the charger on each other, it is important to probe the interplay between the coherent exchange interaction between $A$ and $B$, and the tunable engineered collective  reservoir in contact with both $A$ and $B$. For simplicity, we assume a suitably engineered collective reservoir with $|\mathcal{A}|=|\mathcal{B}|=1$ (see Eq.~(\ref{eq:global_jump_operator}))\footnote{One can also work out the results using $|\mathcal{A}|\neq|\mathcal{B}|\neq 1$ without any qualitative change in them, while the expressions for energy become more cumbersome.}. Defining $\xi_{A(B)}=\Gamma+\gamma_{A(B)}$, note, from Eq.~(\ref{eq:Opeator_evolution_gen_1}),  that the expectation value of an arbitrary observable $\mathcal{O}_A$ defined on $A$ is completely free of the effect of $B$, and is given by
\begin{eqnarray}
\label{eq:two-qubit-A-operator}
\frac{d\left\langle\mathcal{O}_A\right\rangle}{dt}
&=&\text{i}\left\langle[\tilde{H}_0+\tilde{H}_A,\mathcal{O}_A]\right\rangle+\frac{\xi_A}{2} \big[\left\langle a^+[\mathcal{O}_A,a^-]\right\rangle\nonumber\\&&+\left\langle [a^+,\mathcal{O}_A]a^-\right\rangle\big],
\end{eqnarray}
when 
\begin{eqnarray}
    \alpha&=&2n\pi+\frac{\pi}{2},\;
    |J|=\frac{\Gamma}{2},
    \label{eq:onlyAonB}
\end{eqnarray}
where we have assumed $\mathcal{A}$ and  $\mathcal{B}$ to satisfy $\mathcal{A}^*\mathcal{B}=\text{e}^{\text{i}\phi}$, and have defined $\alpha=\theta-\phi$ (see Appendix~\ref{app:reciprocity}). In a similar fashion, the effect of $A$ on an observable $\mathcal{O}_B$ defined on $B$ can be completely removed by fixing 
\begin{eqnarray}
    \alpha&=&2n\pi+\frac{3\pi}{2},\;|J|=\frac{\Gamma}{2},
    \label{eq:onlyBonA}
\end{eqnarray}
where one obtains
\begin{eqnarray}
\frac{d\left\langle\mathcal{O}_B\right\rangle}{dt}
&=&\text{i}\left\langle[\tilde{H}_B,\mathcal{O}_B]\right\rangle+\frac{\xi_B}{2}\big(\left\langle b^+[\mathcal{O}_B,b^-]\right\rangle\nonumber\\&&+\left\langle [b^+,\mathcal{O}_B]b^-\right\rangle\big).
\label{eq:two-qubit-B-operators}
\end{eqnarray} 
See Appendix~\ref{app:reciprocity} for details. Eqs.~(\ref{eq:onlyAonB}) and (\ref{eq:onlyBonA}) imply a complete nonreciprocity (CNR) between $A$ and $B$, where we denote the absence of the effect of $A$ ($B$) on $B$ ($A$) by $\text{CNR}_{A\not\to B}$ ($\text{CNR}_{A\not\leftarrow B}$). On the other hand, $A$ and $B$ affect each other identically if 
\begin{eqnarray}
    \alpha&=&2n\pi,\;|J|=\frac{\Gamma}{2},
    \label{eq:AB_reciprocal}
\end{eqnarray}
representing a complete reciprocity (CR) denoted by $\text{CR}_{A\leftrightarrow B}$. See Fig.~\ref{fig:parameter_space} for a demonstration of the CNR and the CR points on the $(\text{Re}(J),\text{Im}(J))$ plane with $\phi=0$. Note that  one can identify the real axis to be corresponding to the XY interaction~\cite{Lieb1961,Barouch1970,Barouch1971}, while the imaginary axis corresponds to the Dzyaloshinskii-Moriya (DM) interaction~\cite{Anderson1959,Dzyaloshinsky1958,Moriya1960,Moriya1960a} (see also~\cite{Lewenstein2007,Amico2008}).

Our numerical investigation reveals that for large $|F|$, $C^\infty\leq 1/2$ along with $\delta C^\infty\leq 0$ in the majority of the explored parameter space, motivating the investigation for $|F|\ll|J|$, referred to as the \emph{low drive} (LD) limit. In the absence of external driving, let the dynamics of the two-qubit dissipative system be described by 
\begin{eqnarray}
    \frac{d\tilde{\rho}}{dt}=\tilde{\mathcal{L}}_0(\tilde{\rho}),
    \label{eq:zeroth_order}
\end{eqnarray}
with 
\begin{eqnarray}
    \tilde{\mathcal{L}}_0(\tilde{\rho})=-\text{i}\left[H_{AB},\tilde{\rho}\right]+\mathcal{D}(\tilde{\rho}),
\end{eqnarray}
where $\mathcal{D}(.)$ is as given in Eq.~(\ref{eq:cal_D}).  Without loss of generality, we assume that all parameters in the system Hamiltonian and the external drive are scaled by $h$, and hence dimensionless. Considering the external driving represented by $\tilde{H}_0$ as the perturbation, Eq.~(\ref{eq:zeroth_order}) modifies to 
\begin{eqnarray}
        \frac{d\tilde{\rho}}{dt}=\tilde{\mathcal{L}}_0(\tilde{\rho})+|F|\tilde{\mathcal{L}}_F(\tilde{\rho}),
\label{eq:zeroth_order_perturbed}
\end{eqnarray}
where $\tilde{\mathcal{L}}_F(\rho)=-\text{i}\left[\left(a^++a^-\right),\tilde{\rho}\right]$.  Dropping terms with $\mathcal{O}(|F|^{\ell>2})$, i.e., terms with contributions of third order in $|F|$, or higher (see Appendix~\ref{app:perturbation}), one obtains a closed set of coupled linear equations in terms of expectation values of operators as 
\begin{eqnarray}
    \langle b^+b^-\rangle_{2}^\infty&=&-\frac{2}{\xi_B}\text{Re}\left[S^*\langle a^+b^-\rangle_{2}^\infty\right],\nonumber\\
    \langle a^+b^-\rangle_{2}^\infty&=&\frac{2}{\xi_A+\xi_B}\bigg(\text{i}|F|\langle b^-\rangle_{1}^\infty\nonumber\\&&-S\langle a^+a^-\rangle_{2}^\infty-R^*\langle b^+b^-\rangle_{2}^\infty\bigg),\nonumber\\
    \langle a^+a^-\rangle_{2}^\infty&=&\frac{2}{\xi_A}\bigg(\text{Re}\left[\text{i}|F|\langle a^-\rangle_{1}^\infty\right]-\text{Re}\left[R\langle a^+b^-\rangle_{2}^\infty\right]\bigg),\nonumber\\
    \langle b^-\rangle_{1}^\infty&=&-\frac{2S}{\xi_B}\langle a^-\rangle_{1}^\infty,\nonumber\\
    \langle a^-\rangle_{1}^\infty&=&-\frac{2}{\xi_A}\left(R\langle b^-\rangle_{1}^\infty+\text{i}|F|\right),
    \label{eq:two-qubit_linear_equations}
\end{eqnarray}
where we have defined (see discussions following Eq.~(\ref{eq:onlyAonB}), and Appendix~\ref{app:reciprocity})
\begin{eqnarray}
    R&=&\text{i}|J|\text{e}^{\text{i}\theta}+\frac{\Gamma \text{e}^{\text{i}\phi}}{2},\nonumber\\
    S&=&\text{i}|J|\text{e}^{-\text{i}\theta}+\frac{\Gamma \text{e}^{-\text{i}\phi}}{2},
    \label{eq:RS}
\end{eqnarray}
and the subscripts denote the perturbation order, indicating that the non-zero contribution in steady-state battery energy (i.e., $\langle b^+b^-\rangle^\infty$) comes only in second order in $|F|$. Solving for $\langle b^+b^-\rangle$, one obtains 
\begin{eqnarray}
        E^{\infty}&=&\frac{4h|F|^2\left(4|J|^2+\Gamma^2+4|J|\Gamma\sin\alpha\right)}{\left[\xi_A\xi_B+4|J|^2-\Gamma^2\right]^2+16|J|^2\Gamma^2\cos^2\alpha}.
    \label{eq:EB_ss_two_qubit}
\end{eqnarray}
Therefore, with $E^{\max}=h$, and for fixed $|J|,|F|,\Gamma$, and $\xi_{A,B}$, the maximum $C^\infty$ is determined by the $\alpha$ that maximizes $E^\infty$, given by  $\alpha=2n\pi+\pi/2$, $n\in\mathbb{Z}$, when $\xi_A\xi_B+4|J|^2-\Gamma^2\neq 0$. Eq.~(\ref{eq:EB_ss_two_qubit}) with $\Gamma=0$ leads to an $\alpha$-independent $E_{0}^\infty$:
\begin{eqnarray}
        E_{0}^\infty&=&\frac{16h|F|^2|J|^2}{\left(\gamma_A\gamma_B+4|J|^2\right)^2},
        \label{eq:EB0_ss_two_qubit}
\end{eqnarray}
implying the same maximization condition for $\delta C^\infty$ w.r.t. $\alpha$ as in the case of $C^\infty$. Subsequently, we write the following proposition, which establishes the DM interaction with $\theta=2n\pi+\pi/2, n\in\mathbb{Z}$  between the qubit battery and the charger to be the optimal in the LD limit if $\phi=0$, thereby motivating us to restrict $J$ accordingly in our investigations.

\prop{2} \emph{For a single-qubit battery indirectly charged by a  weakly driven single-qubit charger, both $C^\infty$ and $\delta C^\infty$ maximize at $\alpha=2n\pi+\pi/2$, $n\in\mathbb{Z}$, i.e., when the interaction between the charger and the battery is of Dzyaloshinskii-Moriya type, and $\xi_A\xi_B+4|J|^2-\Gamma^2\neq 0$, when all other parameters remain fixed.}

\subsection{High-drive limit  \texorpdfstring{at $\alpha=2n\pi+\pi/2, n\in\mathbb{Z}$}{on the positive DM line}}
\label{subsec:high_drive}

Clearly, for large $|F|$, the optimizations may no longer take place at $\alpha=2n\pi+\pi/2, n\in\mathbb{Z}$. However,  considering the \emph{symmetric limit} $\gamma_A=\gamma_B=\gamma$ (i.e., $\xi_A=\xi_B=\xi=\gamma+\Gamma$) only for this subsection (Sec.~\ref{subsec:high_drive}),  and fixing $\alpha=2n\pi+\pi/2, n\in\mathbb{Z}$, $\langle b^+b^-\rangle^\infty$ can be calculated by solving the following closed set of coupled linear equations
\begin{eqnarray}
    \langle b^{+}b^{-}\rangle^{\infty} &=& -u_+\mathrm{Re}\langle a^{+}b^{-}\rangle/\xi,\nonumber\\
    \langle a^{+}b^{-}\rangle^{\infty} &=& [ 4\Gamma\langle a^{+}a^{-}b^{+}b^{-}\rangle-u_-\langle b^{+}b^{-}\rangle - u_+\langle a^{+}a^{-}\rangle \nonumber\\&&+ 2\mathrm{i}|F|(\langle b^{-}\rangle- 2\langle a^{+}a^{-}b^{-}\rangle)]/2\xi,\nonumber\\
    \langle a^{+}a^{-}\rangle^{\infty} &=&[2|F|\mathrm{Im}\langle a^{+}\rangle-u_-\mathrm{Re}\langle a^{+}b^{-}\rangle]/\xi,\nonumber\\
    \langle a^{+}\rangle^{\infty} &=& [u_-(2\langle a^{+}a^{-}b^{+}\rangle - \langle b^{+}\rangle) \nonumber\\&&- 2\mathrm{i}|F|(1-2\langle a^{+}a^{-}\rangle) ]/\xi,\nonumber\\
    \langle b^{+}\rangle^{\infty} &=& [u_+(2\langle a^{+}b^{+}b^{-}\rangle - \langle a^{+}\rangle)]/\xi,\nonumber\\
    \langle a^{+}a^{-}b^{+}\rangle^{\infty} &=&[2\mathrm{i}|F|(\langle a^{-}b^{+}\rangle - \langle a^{+}b^{+}\rangle) \nonumber\\&&-u_-\langle a^{+}b^{+}b^{-}\rangle]/3\xi,\nonumber\\
    \langle a^{+}b^{+}b^{-}\rangle^{\infty} &=& [2\mathrm{i}|F|(\langle b^{+}b^{-}\rangle- 2\langle a^{+}a^{-}b^{+}b^{-}\rangle) \nonumber\\&&-u_+\langle a^{+}a^{-}b^{+}\rangle]/3\xi,\nonumber\\
    \langle a^{+}b^{+}\rangle^{\infty} &=& \mathrm{i}|F|[\langle b^{+}\rangle-2\langle a^{+}a^{-}b^{+}\rangle]/\xi,\nonumber\\
    \langle a^{+}a^{-}b^{+}b^{-}\rangle^{\infty} &=& |F|\mathrm{Im}\langle a^{+}b^{+}b^{-}\rangle/\xi,
\end{eqnarray}
where we have defined $u_\pm=\Gamma \pm 2|J|$, 
leading to 
\begin{eqnarray}
        E^\infty&=&hP/Q,
        \label{eq:battery_energy_general}
\end{eqnarray}
with 
\begin{eqnarray}
        P&=& 32u_+^{2}|F|^{6} + 8u_+^{2}\left(9\xi^{2}+2|J|u_+\right)|F|^{4}\nonumber\\&&+ 4u_+^{2}\xi^{2}\left(9\xi^{2}-u_+u_-\right)|F|^{2},\nonumber \\
        Q&=&64\left(4\xi^{2}+u_+^{2}\right)|F|^{6}+32\big[10\xi^{4}\nonumber\\&&+u_+(9\Gamma+11|J|)\xi^{2}-u_+^{2}u_-(\Gamma+|J|)\big]|F|^{4}\nonumber\\&&+4\big[27\xi^{6}+ 2u_+(\Gamma+12|J|)\xi^{4}\nonumber\\&&+u_+^{2}u_-(3\Gamma-2|J|)\xi^{2} \big]|F|^{2}\nonumber\\&&+\xi^{2}\left(\xi^{2}-u_+u_-\right)^{2}\left(9\xi^{2}+4J^{2}-\Gamma^{2}\right).
\end{eqnarray}
In  the high-drive (HD) limit $|F|\to\infty$, steady-state battery energy can be obtained using the L'H\^{o}pital's rule as
\begin{eqnarray}
        \lim_{|F|\to\infty}E^\infty&\equiv& E_{\text{HD}}^\infty =\frac{hu_+^2}{2\left(u_+^2+4\xi^2\right)}.
        \label{eq:energy_high_drive_limit}
\end{eqnarray}
Identifying $E^{\max}=h$,
\begin{eqnarray}
    C_{\text{HD}}^\infty=\frac{E_{\mathrm{HD}}^\infty}{E^{\max}}<\frac{1}{2},
\end{eqnarray}
since $\xi> 0$ (see Eq.~(\ref{eq:irreducibility_validity})).

\begin{figure}
    \includegraphics[width=\linewidth]{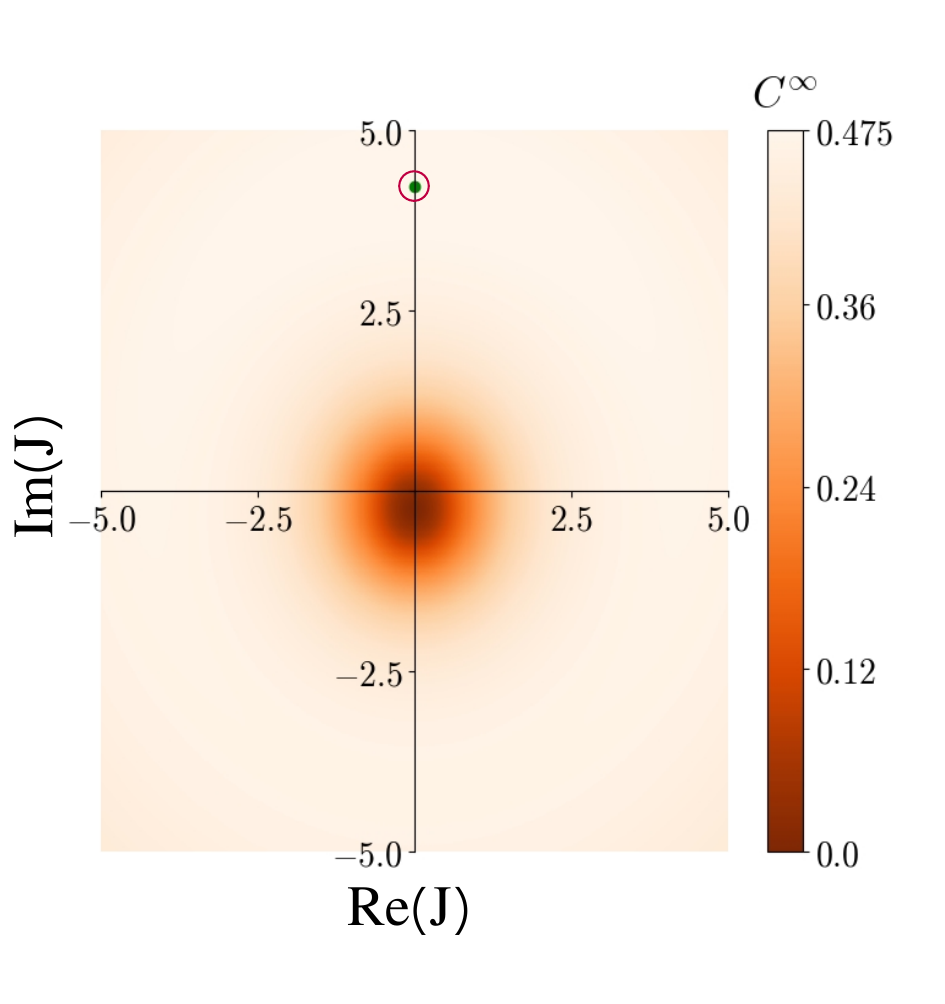}
    \caption{Variation of $C^{\infty}$ of a single-qubit battery as a function of $(\text{Re}(J),\text{Im}(J))$ with a strongly driven ($|F|=2$) single-qubit charger. The values of the other relevant parameters are fixed at $\gamma_A=\gamma_B=0.6$, and $\Gamma=0.5$. The optimal $|J|$ (i.e., $|J|_0$), occurring on the $\alpha=\pi/2$ line, is marked.}
    \label{fig:HD_2qubits}
\end{figure}

\begin{table*}[t]
\caption{Conditions for advantage over the CRF limit at fixed $|J|$ for $\Gamma>0$, with $N_A$ chargers and one battery.}
\label{tab:dC_sign_NA}
\begin{ruledtabular}
\begin{tabular}{lll}
Regime & Condition & Sign of $\delta C^{\infty}$\\
\hline
Weak coupling & $2|J|<\min(\gamma_A/N_A,\gamma_B)$ & $\delta C^{\infty}>0$\\
Intermediate coupling & $\min(\gamma_A/N_A,\gamma_B)<2|J|<\max(\gamma_A/N_A,\gamma_B)$ & $\delta C^{\infty}<0$\\
Strong coupling & $2|J|>\max(\gamma_A/N_A,\gamma_B)$ & $\delta C^{\infty}>0$\\
Boundaries & $2|J|=\gamma_A/N_A$ or $2|J|=\gamma_B$ & $\delta C^{\infty}=0$\\
Graph-adapted symmetric limit & $\gamma_A/N_A=\gamma_B$ (i.e., $N_A=\gamma_A/\gamma_B$) & $\delta C^{\infty}\geq0$ for all $|J|$, zero only at $2|J|=\gamma_B$\\
\end{tabular}
\end{ruledtabular}
\end{table*}

We now focus on the extrema of $C_{\text{HD}}^\infty$ that can be reached by tuning $|J|$ alone. Since $\xi>0$, 
\begin{eqnarray}
        \frac{dC_{\text{HD}}^\infty}{d|J|}= \frac{8u_+\xi^2}{\left(u_+^2+4\xi^2\right)^2},
\end{eqnarray}
obtained from Eq.~(\ref{eq:energy_high_drive_limit}), is strictly positive for all $|J|\geq0$, and vanishes only at the boundary $u_+=0$, i.e., for $\Gamma=0$ and $|J|=0$. Since $C_{\text{HD}}^\infty$ is differentiable w.r.t. $|J|$ and $dC_{\text{HD}}^\infty/d|J|>0$ $\forall |J|\in(0,\infty)$, $C^\infty$ is strictly increasing with $|J|$ in $[0,\infty)$ for fixed $\xi$, and attains the maximum only at $|J|\to\infty$, i.e., $u_+\to\infty$, given by  
\begin{eqnarray}
        C^\infty_{\mathrm{HD,max}}=\left. C_{\text{HD}}^\infty\right|_{u_+\to\infty}=\frac{1}{2}. 
\end{eqnarray}
Also, for $|J|\to 0$, $u_+\to\Gamma$. This, from Eq.~(\ref{eq:energy_high_drive_limit}) and from the fact that $C^\infty$ is strictly increasing for $|J|\in[0,\infty]$, leads to 
\begin{eqnarray}
        C^\infty_{\mathrm{HD,min}}=\left.C_{\text{HD}}^{\infty}\right|_{u_+\to\Gamma} &=& \frac{\Gamma^2}{2\left(\Gamma^2+4\xi^2\right)}
\end{eqnarray}
which is $>0$ as long as $\Gamma>0$. Further, determining $E_{0}^\infty$ using Eq.~(\ref{eq:energy_high_drive_limit}) with $\Gamma=0$, in the limit $|F|\to\infty$, one obtains
\begin{eqnarray}
        \delta C_{\mathrm{HD}}^\infty &=& \frac{\left(\gamma u_+-2|J|\xi\right)\left(\gamma u_++2|J|\xi\right)}{2\left(u_+^{2}+4\xi^{2}\right)\left(|J|^{2}+\gamma^{2}\right)}.
\end{eqnarray}
Note that the sign of $\delta C_{\mathrm{HD}}^\infty$ is determined entirely by the numerator. Therefore, $\delta C_{\mathrm{HD}}^\infty>0$ requires $\Gamma(\gamma - 2|J|)[\Gamma(\gamma + 2|J|) + 4\gamma |J|] > 0$. Since the term within the square brackets is positive, for $\delta C_{\mathrm{HD}}^\infty>0$, $|J|< \gamma/2$. We summarize these results in the following proposition. 

\prop{3} \emph{With $\alpha=2n\pi+\pi/2$ ($n\in\mathbb{Z}$) and $|F|\to\infty$, in the symmetric limit $\gamma_A=\gamma_B=\gamma$, the two-qubit battery-charger setup has the following features. 
\begin{enumerate}
    \item $C^\infty$ is upper bounded by $\tfrac{1}{2}$, which is achieved if and only if $|J|\to\infty$.
    \item For $|J|\to 0$, $C^\infty> 0$ if $\Gamma> 0$, and $C^{\infty}=0$ if $\Gamma=0$.
    \item An advantage over the CRF limit $(\delta C^\infty>0)$ is obtained only when $0\leq |J|<\gamma/2$
\end{enumerate}}

\noindent These results indicate that the values of $|J|$ to achieve  $C_{\mathrm{HD,max}}^{\infty}$,  and to engineer an advantageous collective bath resulting in $\delta C_{\mathrm{HD}}^\infty>0$ do not coincide in the HD limit.  It is also important to stress that the above results are true \emph{only} for $\alpha=2n\pi+\pi/2, n\in\mathbb{Z}$, further implying the sufficiency of investigating the LD limit alone. The maxima of $C^\infty$ and $\delta C^\infty$ w.r.t $\alpha$ for arbitrary values of $F$ and $J$ may still occur at a value of $\alpha\neq \pi/2$. However, even for high $|F|$, there exist points in parameter space where the optimal $C^\infty$ takes place on the $\alpha=2n\pi+\pi/2$ ($n\in\mathbb{Z}$) line, as demonstrated below.

Fig.~\ref{fig:HD_2qubits} depicts $C^\infty$ on the $(\mathrm{Re}(J),\mathrm{Im}(J))$ plane for a strongly driven single-qubit charger. The capacity is maximized on the  $\alpha=\pi/2$ line, while the maximum remains below the bound $1/2$ (see
Proposition~3). The choice of a finite $|F|$ results in the  maximum occurring at a finite
coupling rather than at $|J|\to\infty$, and $C^\infty$ remains close to its maximum
over a broad range of large $|J|$. It is worthwhile to point out here that the bound $C^{\infty}\leq 1/2$ stems from the saturability of the qubit, and has no counterpart in the oscillator models, where the stored energy grows with the drive strength without bound~\cite{Zafar2026,Khan2026,Khan2026arX}.

\subsection{Low-drive limit  \texorpdfstring{at $\alpha=2n\pi+\pi/2, n\in\mathbb{Z}$}{on the positive DM line}}
\label{subsec:low_drive}

We now explore the LD limit. We start with Eq.~(\ref{eq:EB_ss_two_qubit}), and determine $C^\infty$ in the LD limit at $\alpha=2n\pi+\pi/2, n\in\mathbb{Z}$ as 
\begin{eqnarray}
    C_{\mathrm{LD}}^{\infty}= \frac{4|F|^{2}u_+^{2}}{\left(\xi_A\xi_B-u_+u_-\right)^{2}}.
    \label{eq:LD_twoqubit_capacity_DMline}
\end{eqnarray}
Determining the physical (non-negative) root from $dC_{\mathrm{LD}}^\infty/d|J|=0$, the maximum of $C^\infty_{\mathrm{LD}}$ is found to occur at 
\begin{eqnarray}
        |J|_0=\frac{\sqrt{\xi_A\xi_B}-\Gamma}{2},
\end{eqnarray}
where positivity of $|J|_0$ is ensured by at least one of $\gamma_{A,B}>0$. At $|J|=|J|_0$, 
\begin{eqnarray}
        C^\infty_{\text{LD,max}}=\left.C_{\mathrm{LD}}^\infty\right|_{|J|=|J|_0}=\frac{|F|^2}{4|J|_0^2}. 
\end{eqnarray}
For validity of the  perturbative arguments, $|F|^2/|J|_0^2<1$, thereby providing an upper bound on maximum $C^\infty_{\mathrm{LD}}$ as  $C_{\mathrm{LD,max}}^\infty<1/4$. Note also that the maximum $C^\infty_{\mathrm{LD}}$ is obtained at the $\text{CNR}_{A\not\leftarrow B}$ point iff  $\xi_A\xi_B=4\Gamma^2$, which can be proved by setting $|J|_0$ equal to the $\text{CNR}_{A\not\leftarrow B}$ point, i.e., $|J|_0=\Gamma/2$.  In the symmetric limit with $\gamma_A=\gamma_B=\gamma$, this reduces to $\Gamma=\gamma$.

From Eqs.~(\ref{eq:EB0_ss_two_qubit}) and (\ref{eq:LD_twoqubit_capacity_DMline}), the CRF capacity is maximized at $2|J|=\sqrt{\gamma_A\gamma_B}$, with $C_{0,\max}^{\infty}=|F|^2/(\gamma_A\gamma_B)$. Note that $\xi_A\xi_B\geq (\Gamma+\sqrt{\gamma_A\gamma_B})^2$ (since $\gamma_A+\gamma_B\geq 2\sqrt{\gamma_A\gamma_B}$), leading to $2|J|_0\geq \sqrt{\gamma_A\gamma_B}$, and subsequently $C_{\text{LD,max}}^{\infty}\leq C^{\infty}_{0,\max}$, with equality occurring only at $\gamma_A=\gamma_B$. Therefore, once the coupling is tuned optimally, the shared reservoir does not increase the maximum low-drive capacity. It can only shift the position of the optimum on the $(|J|,\alpha)$ plane.

\begin{figure*}
    \includegraphics[width=\linewidth]{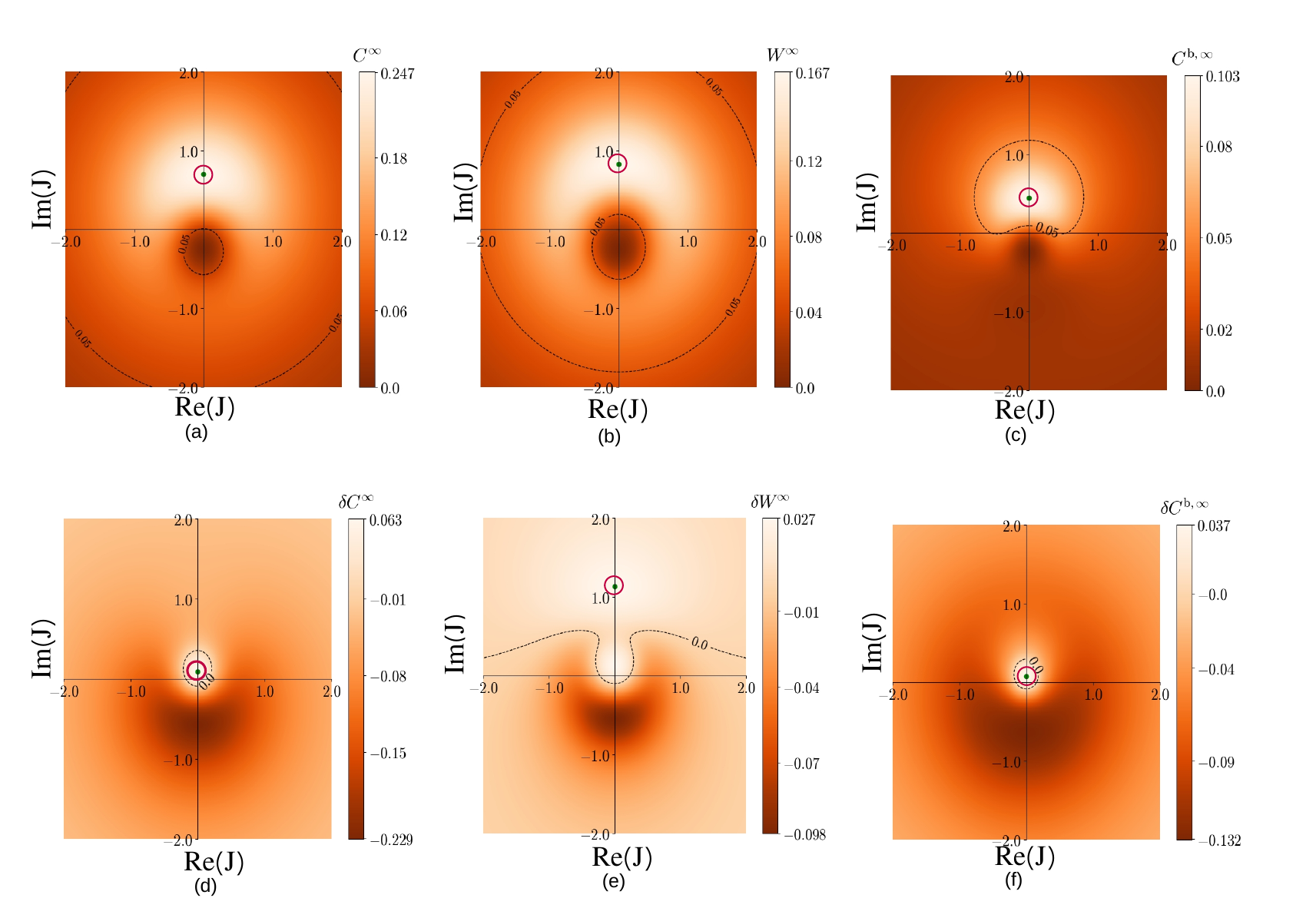}
    \caption{Variation of capacity ((a) $C^{\infty}$, (b) $W^\infty$), advantage over the CRF limit ((d) $\delta C^\infty$, (e) $\delta W^\infty$), bound capacity ((c) $C^{\rm{b}}$), and the advantage in bound capacity ((f) $\delta C^{\rm{b},\infty}$) of a single-qubit battery charger through a weakly-driven single-qubit charger as a function of ($\text{Re}(J),\text{Im}(J)$). The values of the other relevant parameters are fixed at $\gamma_A=\gamma_B=0.6$, $\Gamma=0.5$, and $|F|=0.5$. The points on the $\alpha=\pi/2$ line mark the values of $|J|_0$ for the corresponding quantities.}
    \label{fig:LD_2qubits}
\end{figure*}

By definition,
\begin{eqnarray}
    \delta C^\infty &=& 4|F|^{2}\left[\frac{u_+^{2}}{\left(\xi_A\xi_B-u_+u_-\right)^{2}}-\frac{4|J|^2}{\left(\gamma_A\gamma_B+4|J|^2\right)^2}\right],\nonumber\\
\end{eqnarray}
is positive when 
\begin{eqnarray}
    \frac{u_+}{2|J|}>\frac{\xi_A\xi_B-u_+u_-}{\gamma_A\gamma_B+4|J|^2},
\end{eqnarray}
which, for $\Gamma>0$, simplifies to 
\begin{eqnarray}
    (2|J|-\gamma_A)(2|J|-\gamma_B)>0. 
\end{eqnarray}
The different possibilities arising out of this are summarized in Table~\ref{tab:dC_sign}. Further, note that $\min(\gamma_A,\gamma_B)\leq 2|J|_0\leq \max(\gamma_A,\gamma_B)$, implying that the signs of $(2|J|_0-\gamma_A)$ and $(2|J|_0-\gamma_B)$ are opposite. Therefore, $\delta C^\infty<0$ at $|J|=|J|_0$.

From the above calculations, the fourfold enhancement reported for bosonic nonreciprocal
batteries~\cite{Ahmadi2024} can be recovered in the LD limit. Comparing
Eqs.~(\ref{eq:EB0_ss_two_qubit}) and (\ref{eq:LD_twoqubit_capacity_DMline}) at $|J|=\Gamma/2$,
\begin{eqnarray}
r_1=\frac{\left.C^\infty_{\rm LD}\right|_{|J|=\Gamma/2}}{\left.C^\infty_0\right|_{|J|=\Gamma/2}}
=4\left[\frac{\gamma_A\gamma_B+\Gamma^2}{(\Gamma+\gamma_A)(\Gamma+\gamma_B)}\right]^2.
\end{eqnarray}
Besides, at the $\text{CNR}_{A\not\leftarrow B}$ point, Eq.~(\ref{eq:two-qubit_linear_equations}) yields $\langle a^-\rangle^\infty_1=-2\text{i}|F|/\xi_A$, and $\langle a^+a^-\rangle_2^{\infty}=2\rm{Re}\left[\text{i}|F|\langle a^-\rangle_1^{\infty}\right]/\xi_A$, leading to a steady state charger capacity $C^\infty_A=4|F|^2/\xi_A^2$. Eq.~(\ref{eq:LD_twoqubit_capacity_DMline}) at $|J|=\Gamma/2$ yields $C_B^\infty=16|F|^2\Gamma^2/\xi_A^2\xi_B^2$, leading to the battery-to-charger capacity ratio
\begin{eqnarray}
    r_2=\frac{\left.C^\infty_{\rm{LD}}\right|_{|J|=\Gamma/2}}{C^\infty_A}=\frac{4\Gamma^2}{\xi_B^2}=\frac{4}{(1+\gamma_B/\Gamma)^2},
\end{eqnarray}
coinciding with that of
Ref.~\cite{Ahmadi2024}. Both $r_1$ and $r_2$ are strictly smaller than $4$ for finite
dissipation rates, and approach $4$ only in limiting regimes
($\Gamma\gg\gamma_{A,B}$, $\Gamma\ll\gamma_{A,B}$, or $\gamma_B\to0$). For $r_2$,
this is consistent with the analysis of the bosonic model of
Ref.~\cite{Ahmadi2024} given in Ref.~\cite{Khan2026arX}, where the fourfold
value was likewise shown to require $\gamma_B/\Gamma\to0$\footnote{Note that
Ref.~\cite{Khan2026arX} subsequently argues for a cascaded master equation
without the reservoir-induced local damping, obtaining a sixteenfold
advantage. While our results throughout this work pertain to the master equation of
Ref.~\cite{Metelmann2015}, they are extendable to the proposals of~\cite{Khan2026arX}.}. Moreover, $r_1>1$ if and only if
$(\Gamma-\gamma_A)(\Gamma-\gamma_B)>0$, which is the condition of Table~\ref{tab:dC_sign} at
$2|J|=\Gamma$. In the symmetric limit, $r_1$ attains its minimum
$r_1=1$ at $\Gamma=\gamma$, precisely where the $\text{CNR}_{A\not\leftarrow B}$ point coincides with
the optimal coupling $|J|_0$. The fourfold advantage is thus an advantage at equal coupling far from the optimum. Once each configuration is operated at its own optimal coupling, the shared reservoir yields no enhancement,
$C^\infty_{\rm LD,max}\le C^\infty_{0,\rm max}$.

These results are summarized in the following proposition.  

\prop{4} \emph{With $\alpha=2n\pi+\pi/2$ ($n\in\mathbb{Z}$), in the low-drive limit, the two-qubit battery-charger setup has the following features. 
\begin{enumerate}
    \item $C^{\infty}$ is maximized at $|J|_0=\left(\sqrt{\xi_A\xi_B}-\Gamma\right)/2$, but without any advantage over the collective-reservoir-free limit.
    \item The maximum value of $C^\infty$ is upper bounded by $\tfrac{1}{4}$. 
    \item With all other parameters fixed, and for $\Gamma
    >0$, $\delta C^{\infty}<0$ as long as $\min(\gamma_A,\gamma_B)< 2|J|< \max(\gamma_A,\gamma_B)$. 
    \item The shared reservoir does not improve the optimal capacity. Instead, it relocates the optimum, and expands the range of $|J|$ over which the charging is efficient. 
\end{enumerate}}

The steady-state performance of the single-qubit battery on the $(\text{Re}(J), \text{Im}(J))$ plane in the LD limit is demonstrated in Fig.~\ref{fig:LD_2qubits}. Figs.~\ref{fig:LD_2qubits}(a)--(c) depict $C^\infty$, $W^\infty$, and the
bound capacity $C^{\mathrm{b},\infty}$, while
Figs.~\ref{fig:LD_2qubits}(d)--(f) depict the corresponding advantages
$\delta C^\infty$, $\delta W^\infty$, and $\delta C^{\mathrm{b},\infty}$. In
agreement with Proposition~2, both $C^\infty$ and $W^\infty$ are maximized for the DM interaction between the battery and the charger, with the maximum $C^\infty$ remaining below the bound $1/4$ (see Proposition~4). The ergotropy is maximized at a slightly larger coupling than the energy storage capacity, while the bound capacity peaks at a comparably smaller coupling, implying that the coupling that
maximizes the stored energy does not maximize the extractable work, and the
fraction of stored energy locked in the battery is largest below the capacity
optimum. All three quantities vanish at the $\text{CNR}_{A\not\to B}$ point (see Fig.~\ref{fig:parameter_space}), where the battery is decoupled from the charger.

The advantage $\delta C^\infty$ (see Fig.~\ref{fig:LD_2qubits}(d)) is only weakly positive only
in a small region around the origin, i.e., at weak
coupling, and is negative elsewhere, most strongly for $\text{Im}(J)<0$ (see Table~\ref{tab:dC_sign}). In contrast, the shared reservoir
enhances the extractable work over a broad region corresponding to $\text{Im}(J)>0$ (see Fig.~\ref{fig:LD_2qubits}(e)), with the maximum reaching at couplings where it reduces the capacity.
Consistently, $\delta C^{\mathrm{b},\infty}<0$ over most of the plane, implying that the \emph{the shared reservoir reduces the fraction of the stored energy that is locked in the battery, even where it reduces the stored energy itself}, clearly underlining the advantage of the engineered reservoir.

\begin{figure}
    \centering
    \includegraphics[width=\linewidth]{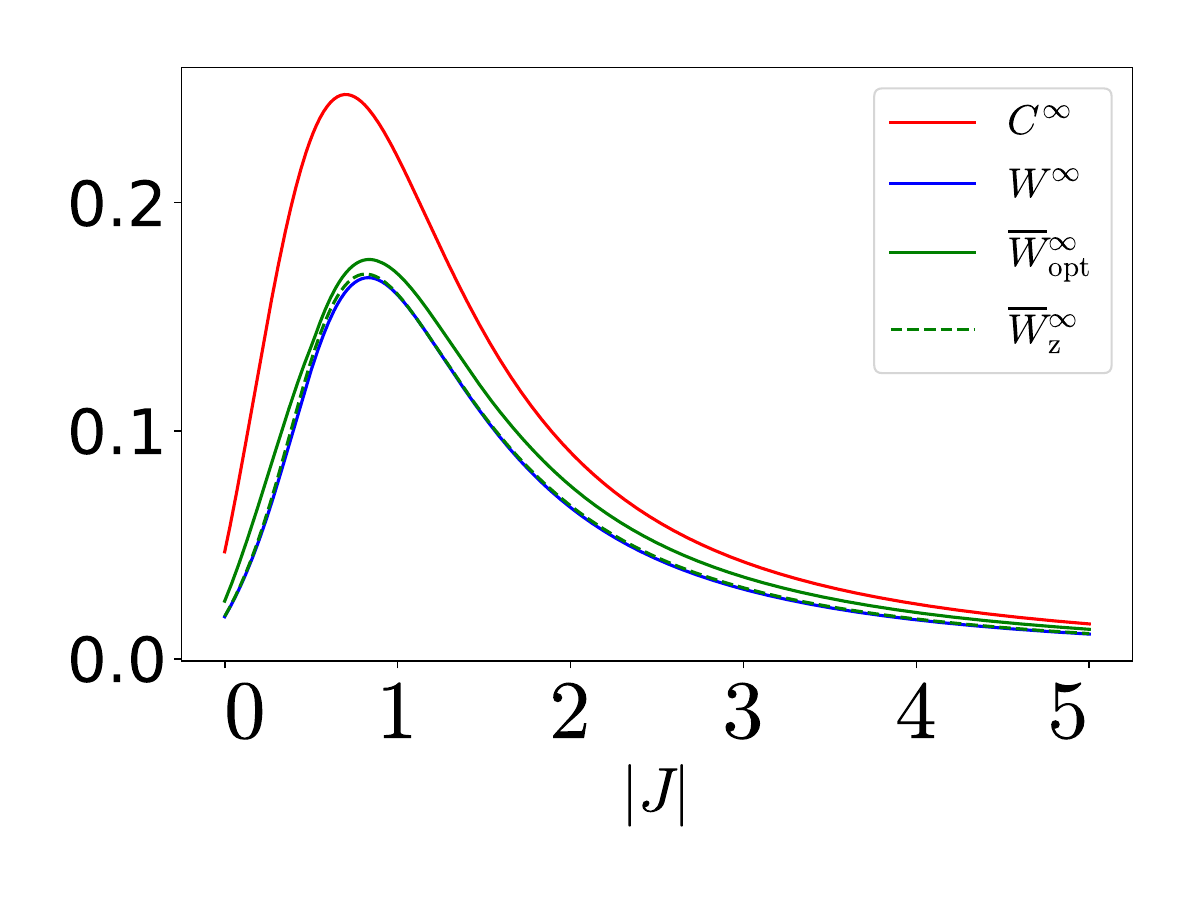}
    \caption{Variations of $C^\infty$, $W^\infty$, $\overline{W}^\infty_z$, and  $\overline{W}^\infty_{\rm{opt}}$ as functions of $|J|$ with $\alpha=\pi/2$. Other relevant parameters are fixed at $\gamma_A=\gamma_B=0.6$, $\Gamma=0.5$, and $|F|=0.5$.}
    \label{fig:bound_capacity}
\end{figure}

To further probe the interplay between the stored and the extractable energy,
in Fig.~\ref{fig:bound_capacity} we plot $C^\infty$, $W^\infty$, and the daemonic
ergotropy as functions of $|J|$ on the $\alpha=\pi/2$ line, where
$\overline{W}^\infty_{\rm opt}$ is maximized over all projective measurements on
the charger, and $\overline{W}^\infty_z$ corresponds to a measurement in the
computational basis. Starting at a low value for $|J|\to 0$, all quantities maximize at intermediate couplings, and decay for large $|J|$, while the capacity peaking at slightly lower coupling   compared to $W^\infty_z$ and
$\overline{W}^\infty_{\rm opt}$.  Therefore, the bound capacity $C^{\mathrm b,\infty}$, is largest
below the maximum capacity of stored energy, and decreases with increasing $|J|$. Owing to the convexity of ergotropy, $\overline{W}^\infty_{\rm opt}\ge\overline{W}^\infty\ge W^\infty$ for all $|J|$, so that \emph{the bound capacity can always be reduced by
extracting work via measurements on the charger}. The gain is, however, measurement-dependent, and low. The computational-basis measurement offers almost no advantage over $W^\infty$,
while the optimal measurement increases the extractable work slightly near the peak. 

It is important to note that the existence of the bound capacity here is a consequence of the finite-dimensional DoFs, and is in contrast with the linearly driven oscillator batteries initialized in vacuum, for which the entire energy stored in the steady state can be extracted~\cite{Khan2026arX,Guo2026PhaseTunable,Zafar2026}. In these latter cases, ergotropy and stored energy may differ under two-photon driving~\cite{Xu2025arX}, or with squeezed and thermal reservoirs~\cite{Liu2026}. Note also that $C^\infty$ vanishes on the negative imaginary axis at $|J|=\Gamma/2$, i.e., at the $\text{CNR}_{A\not\to B}$
point [Eq.~(\ref{eq:onlyBonA})], where the battery is decoupled from the charger, as from Eq.~(\ref{eq:EB_ss_two_qubit}), the numerator becomes
$(2|J|-\Gamma)^2=0$ at $\alpha=3\pi/2$.

\section{Weakly-driven charger-battery setup on a star network}
\label{sec:multiqubit_charger}

We now consider a complete bipartite graph representing the connectivity between the qubits (see Appendix~\ref{app:reciprocity}), focussing specifically on a star network of qubits.  Here,  each of the qubits $i\in A$ is connected to each of  the qubits $j\in B$ via the coherent exchange interaction of strength $J_{i,j}$, but no intra-species interactions exist (i.e., $g_{i,i^\prime}=0$ $\forall\langle i,i^\prime\rangle\in G$, and $g_{j,j^\prime}=0$ $\forall\langle j,j^\prime\rangle\in G$). For simplicity, we assume $J_{i,j}=J$ $\forall\langle i,j\rangle\in G$, implying $\theta_{i,j}=\theta$ $\forall\langle i,j\rangle\in G$. Further, we assume that the system is operating in the LD limit for all qubits in $A$, and $\mathcal{A}_i=\mathcal{A}\forall i\in A$, $\mathcal{B}_j=\mathcal{B}\forall j\in B$, with $\mathcal{A}^*\mathcal{B}=\text{e}^{\text{i}\phi}$.

\subsection{Single central battery qubit}
\label{subsec:star_charger}

\begin{figure}
    \includegraphics[width=0.8\linewidth]{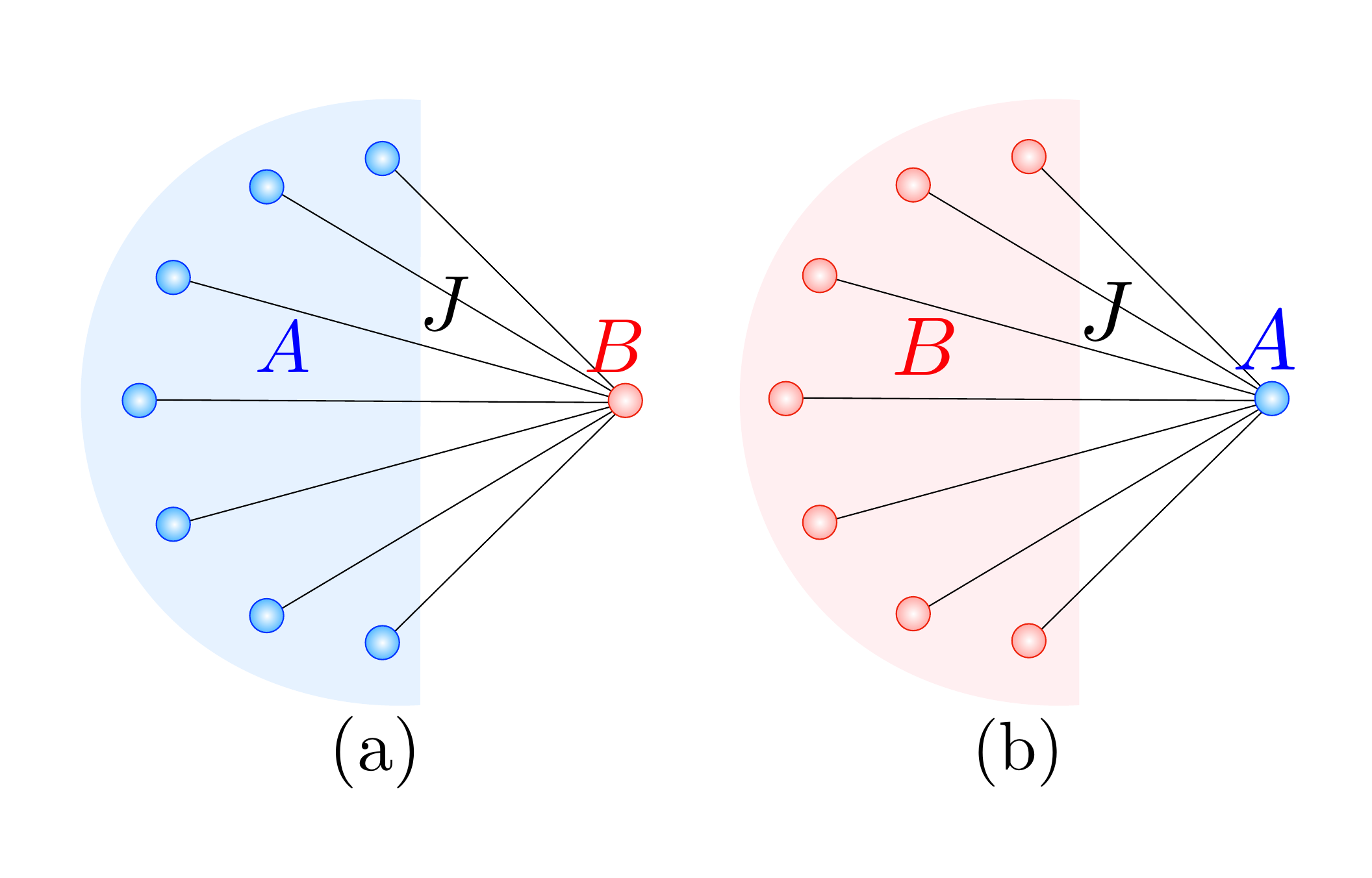}
    \caption{\textbf{Batteries and chargers arranged in a star network.} (a) Here, the central qubit plays the role of the battery, and the peripheral qubits act as the externally driven chargers. (b) The charger is placed at the center of the network, with the batteries arranged around it on the periphery.}
    \label{fig:star_charger}
\end{figure}

\begin{figure}
    \centering
    \includegraphics[width=\linewidth]{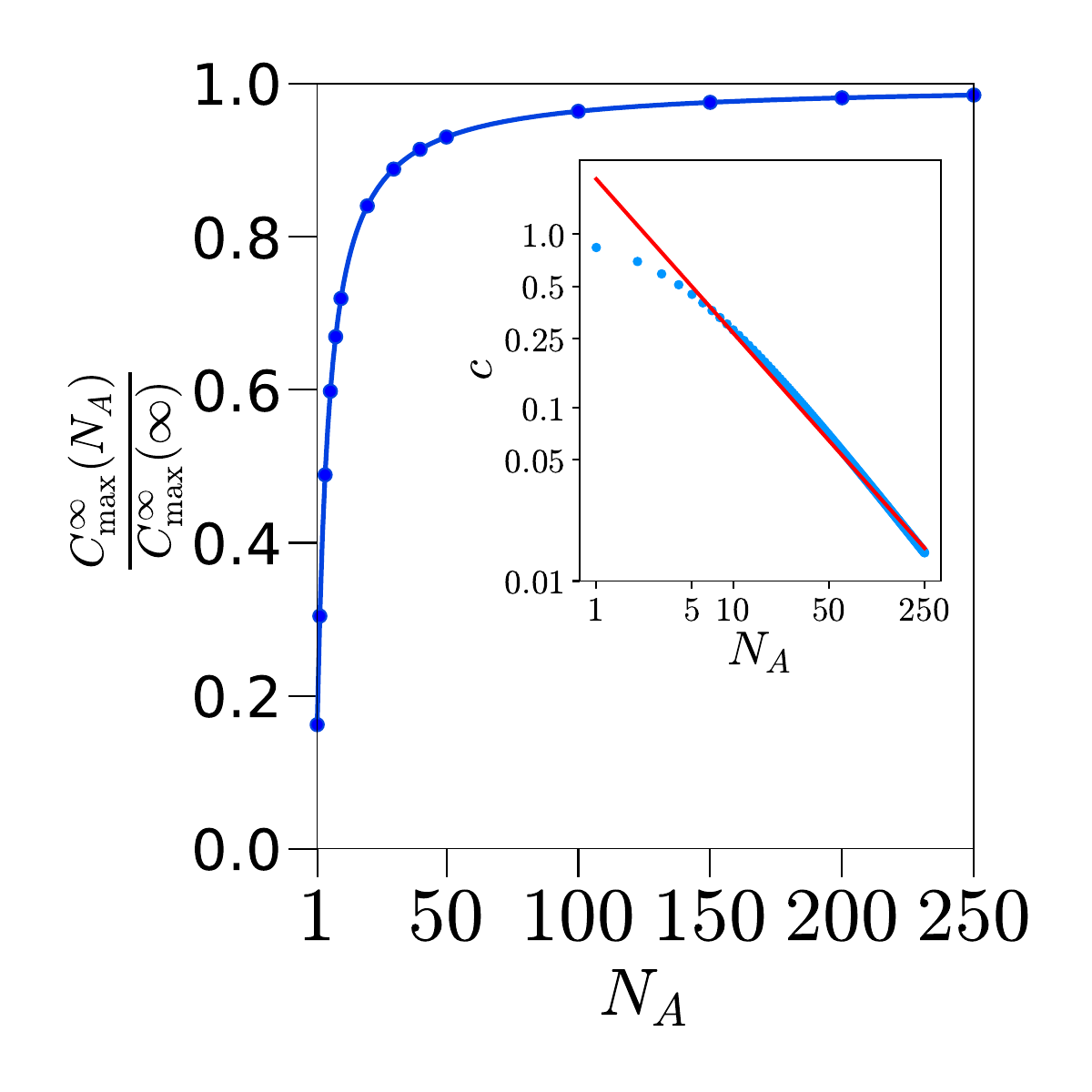}
    \caption{Variation of $C^{\infty}_{\max}(N_A)/C^{\infty}_{\max}(\infty)$ against $N_A$. (Inset) Log-log plot of $c$ with $N_A$. The relevant parameters are fixed at $\gamma_A=\gamma_B=0.6$,  $\Gamma=0.5$, and $|F|=0.5$.}
    \label{fig:capacity_theory}
\end{figure}

We first consider a single qubit at the center of the star network (see Fig.~\ref{fig:star_charger}(a)) posing as the battery, while the peripheral qubits are the externally driven chargers. We adopt the perturbative treatment of the field term, define $\obj{a}^{\pm}=\sum_{i}a_i^{\pm}$, and drop $\mathcal{O}\left(|F|^{\ell>2}\right)$ terms (see Appendix~\ref{app:perturbation}) to obtain the following closed set of coupled linear equations: 
\begin{eqnarray}
    \langle b^+b^-\rangle_{2}^\infty&=&-\frac{2}{\xi_B}\text{Re}\left[S^*\langle \obj{a}^+b^-\rangle_{2}^\infty\right],\nonumber\\
    \langle \obj{a}^+b^-\rangle_{2}^\infty&=&-\frac{2}{N_A\chi_A+\xi_B}\bigg(\text{i}|F|N_A\langle b^-\rangle_{1}^\infty\nonumber\\&&-S\langle \obj{a}^+\obj{a}^-\rangle_{2}^\infty-R^*N_A\langle b^+b^-\rangle_{2}^\infty\bigg),\nonumber\\
    \langle \obj{a}^+\obj{a}^-\rangle_{2}^\infty&=&\frac{2}{\chi_A}\bigg(\text{Re}\left[\text{i}|F|\langle \obj{a}^-\rangle_{1}^\infty\right]-\text{Re}\left[R\langle \obj{a}^+b^-\rangle_{2}^\infty\right]\bigg),\nonumber\\
    \langle b^-\rangle_{1}^\infty&=&-\frac{2S}{\xi_B}\langle \obj{a}^-\rangle_{1}^\infty,\nonumber\\
    \langle \obj{a}^-\rangle_{1}^\infty&=&-\frac{2}{\chi_A}\left(R\langle b^-\rangle_{1}^\infty+\text{i}|F|\right),
\end{eqnarray}
with $R$ and $S$ defined in Eq.~(\ref{eq:RS}).  Solving these equations leads to 
\begin{eqnarray}
    E^\infty &=& \frac{4h|F|^2\left(4|J|^2+\Gamma^2+4|J|\Gamma\sin\alpha\right)}{\left(\chi_A\xi_B+4|J|^2-\Gamma^2\right)^2+16|J|^2\Gamma^2\cos^2\alpha},
    \label{eq:steady_state_energy_NA_chargers}
\end{eqnarray}  
where we have defined $\chi_{A(B)}=\Gamma+\left(\gamma_{A(B)}/N_{A(B)}\right)$. 
Therefore, identifying $E^{\max}=h$, $C^\infty$ is maximized for $\alpha=2n\pi+\pi/2$, $n\in\mathbb{Z}$, similar to the two-qubit battery-charger setup. Note that $\chi_{A(B)}=\xi_{A(B)}$ for  $N_{A(B)}=1$. Using the same notation as in  Sec.~\ref{subsec:low_drive}, from  Eq.~(\ref{eq:steady_state_energy_NA_chargers}), $C^\infty$ at $\alpha=2n\pi+\pi/2$ ($n\in\mathbb{Z}$) becomes  
\begin{eqnarray}
    C^{\infty}= \frac{4|F|^{2}u_+^{2}}{\left(\chi_A\xi_B-u_+u_-\right)^{2}},
    \label{eq:Cmax_1}
\end{eqnarray}
which is maximized at $|J|=|J|_0=(\sqrt{\chi_A\xi_B}-\Gamma)/2$, having a value 
\begin{eqnarray}
    C_{\max}^\infty(N_A) = \frac{|F|^2}{[D(N_A)]^2},
\end{eqnarray}
where we define $D(N_A)=T(N_A)-\Gamma$, and  $T(N_A)=\sqrt{\chi_A\xi_B}$ for brevity, with $|J|_0=D(N_A)/2$. Observe that $T(N_A)$, and subsequently  $D(N_A)$, are strictly decreasing functions of $N_A\in[1,\infty)$, with $T(N_A)>T(\infty)=\sqrt{\Gamma\xi_B}\geq \Gamma$, since $\gamma_B\geq 0$. This implies that for any two consecutive integers $N_A$ and $N_A+1$, $D(N_A)>D(N_A+1)\geq 0$, which, in turn, implies $C_{\max}^\infty(N_A)$ to be a strictly increasing function of $N_A$ (see Fig.~\ref{fig:capacity_theory}).

Let us now assume that $\Gamma>0$. Since $C_{\max}^\infty(N_A)$ is strictly increasing with $N_A$, and since $D(N_A)> D(\infty)\geq 0$, 
\begin{eqnarray}
        C_{\max}^\infty(N_A)<C_{\max}^\infty(\infty)=\frac{|F|^2}{\left(\sqrt{\Gamma\xi_B}-\Gamma\right)^2}\; \forall N_A, 
\end{eqnarray}
providing the saturation value. We are specifically interested in $C_{\max}^\infty(N_A)/C_{\max}^\infty(\infty)=[D(\infty)/D(N_A)]^2$, which allows one to quantify the relative shortfall of reaching the saturation value as 
\begin{eqnarray}
        c&=&1-\frac{C_{\max}^\infty(N_A)}{C_{\max}^\infty(\infty)},\nonumber\\&=&1-\left[\frac{D(\infty)}{D(N_A)}\right]^2,\nonumber\\&\approx& \frac{2[D(N_A)-D(\infty)]}{D(\infty)},
\end{eqnarray}
when $D(N_A)\to D(\infty)$, i.e., when $N_A\to\infty$. Note further that 
\begin{eqnarray}
        D(N_A)-D(\infty)&=&T(N_A)-T(\infty),\nonumber\\&=&\frac{\gamma_A\xi_B}{N_A\left[T(N_A)+T(\infty)\right]},\nonumber\\&>&0,
\end{eqnarray}
implying 
\begin{eqnarray}
        N_A\left[D(N_A)-D(\infty)\right]\to\gamma_A\xi_B/2T(\infty)
\end{eqnarray}
when $N_A\to\infty$. Combining, one obtains
\begin{eqnarray}
        c&\approx& \frac{\gamma_A\xi_B}{N_AT(\infty)D(\infty)}= \frac{\gamma_A\xi_B}{N_A\sqrt{\Gamma\xi_B}\left(\sqrt{\Gamma\xi_B}-\Gamma\right)},
\end{eqnarray}
implying that for large $N_A$, $C_{\max}^{\infty}(N_A)$ approaches $C_{\max}^{\infty}(\infty)$ from below as $\sim N_A^{-1}$ for $\Gamma>0$ (see inset of Fig.~\ref{fig:capacity_theory}). In contrast, for $\Gamma=0$, $T(N_A)=\sqrt{\gamma_A\gamma_B/N_A}$,leading to 
\begin{eqnarray}
        C_{0,\max}^\infty(N_A)=\frac{N_A|F|^2}{\gamma_A\gamma_B}.
\end{eqnarray}

We now consider the advantage $\delta C^\infty$ over the CRF limit. Proceeding as in Sec.~\ref{subsec:low_drive}, the condition for $\delta C^\infty>0$ can be obtained using Eq.~(\ref{eq:Cmax_1}) as 
\begin{eqnarray}
    \frac{u_+}{2|J|}>\frac{N_A(\chi_A\xi_B-u_+u_-)}{\gamma_A\gamma_B+4N_A|J|^2}.
\end{eqnarray}
Similar arguments as in Sec.~\ref{subsec:low_drive} provide the conditions on $|J|$ for achieving $\delta C^\infty>0$ as given in Table~\ref{tab:dC_sign_NA} (which reduces to the conditions given in Table~\ref{tab:dC_sign} for $N_A=1$), and establishes  $\min(\gamma_A/N_A,\gamma_B)\leq 2|J|_0\leq \max(\gamma_A/N_A,\gamma_B)$, leading to $\delta C^\infty<0$ at $|J|=|J|_0$.

\begin{figure*}
    \centering
    \includegraphics[width=\linewidth]{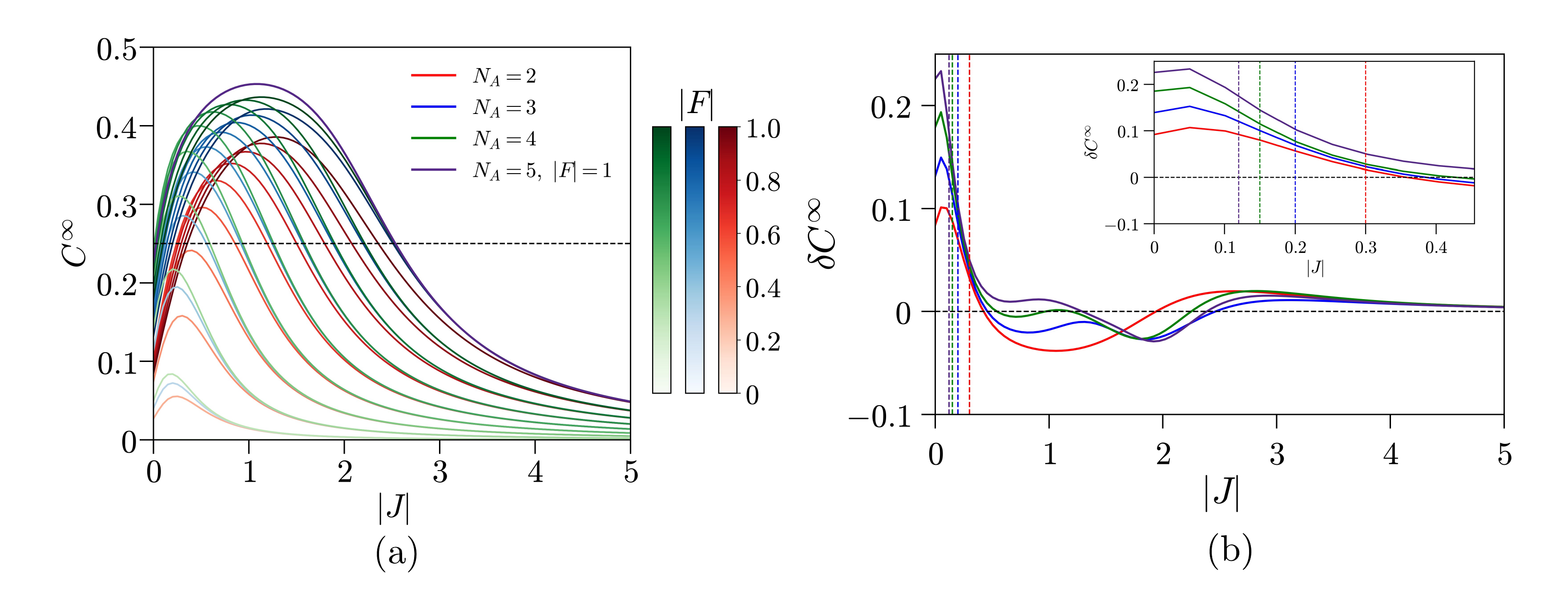}
    \caption{Variations of (a) $C^\infty$ and (b) $\delta C^{\infty}$ as functions of $|J|$ $\left(0\leq |J|\leq 5h\times 10^{-2}\right)$ for different values of $|F|$ ($0\leq |F|\leq 1$) and $N_A\leq 4$, with $h=10^2$, $\gamma_A=\gamma_B=0.6$, $\Gamma=0.5$. (Inset of (b)) Variation of $\delta C^{\infty}$ as function of $|J|$ in the range $0\leq |J|\leq h\times 10^{-2}$. The vertical lines correspond to $\gamma_A/N_A$.}
    \label{fig:HD_star}
\end{figure*}

These results corresponding to weak driving can be summarized in the following proposition. 

\prop{5} \emph{For $N_A$ qubits as chargers and $1$ qubit as battery arranged in a star network with the battery being the central qubit, with fixed $|F|\ll |J|$, $C^\infty$ exhibits the following features at $\alpha=2n\pi+\pi/2$ ($n\in\mathbb{Z}$).
\begin{enumerate}
    \item $C^{\infty}$ is maximized at $|J|_0=\left(\sqrt{\chi_A\xi_B}-\Gamma\right)/2$, but without any advantage over the collective-reservoir-free limit.  
    \item With $N_A\to\infty$ and $\Gamma>0$,  the maximum $C^{\infty}$ saturates to $|F|^2/\left(\sqrt{\Gamma\xi_B}-\Gamma\right)^2$ as $\sim N_A^{-1}$ from below. However, for $\Gamma=0$, the maximum $C^{\infty}$ grows as $\sim N_A$.
    \item With all other parameters fixed, and for $\Gamma
    >0$, $\delta C^{\infty}<0$ as long as $\min(\gamma_A/N_A,\gamma_B)< 2|J|< \max(\gamma_A/N_A,\gamma_B)$. 
\end{enumerate}
}

\subsubsection{Beyond the perturbative regime}
\label{subsubsec:star_charger_strong_driving}

We now explore beyond the weak driving limit and check if the results corresponding to charging a qubit with a single-qubit charger at the HD limit can be qualitatively obtained for $N_A$ chargers also when the weak driving condition is relaxed. To keep the form of the Lindblad operators justified, we confine ourselves to the region $|F|,|J|\lesssim 10^{-2}h$, where $|F|\ll |J|$ ($|F|\approx |J|$) corresponding to the LD (intermediate drive) limit (note that similar parameter space is used in Fig.~\ref{fig:HD_2qubits}). Our numerical investigation reveals the optimal value of $\alpha$ to be $2n\pi+\pi/2$ ($n\in\mathbb{Z}$) within this range of parameters, implying the investigation over the $\alpha=\pi/2$ line to be sufficient.

We now go beyond the perturbative regime, and consider drive strengths comparable to the coherent coupling, restricting to  $|F|,|J|\lesssim10^{-2}h$ with $h=10^2$, so that the local form of the jump operators in Eq.~(\ref{eq:global_jump_operator}) remains justified. Our numerical investigations  find that within this parameter range, the optimum of $C^\infty$ lies
on the line $\alpha=2n\pi+\pi/2$ ($n\in\mathbb{Z}$). Therefore, it suffices to fix $\alpha=\pi/2$, which we do for the rest of the analysis. Fig.~\ref{fig:HD_star}(a) depicts $C^\infty$ as a function of $|J|$
for $N_A=2,3,4$ with discrete values of $|F|$ in the range  $0\le|F|\le1$, and for $N_A=5$ with $|F|=1$, where
$\gamma_A=\gamma_B=0.6$ and $\Gamma=0.5$. For weak drive, $C^{\infty}$ peaks at
small $|J|$, close to the low-drive optima $|J|_0$
predicted for $N_A=2,3,4$ (see Proposition~5), which lie below the $\mathrm{CNR}_{A\not\leftarrow B}$  point $|J|=\Gamma/2=0.25$. With increasing $|F|$, the maximum grows well beyond
the low-drive bound $1/4$ and shifts towards larger $|J|$, while $C^\infty$ further decays at large $|J|$ where
the setup re-enters the low-drive regime. Throughout the explored parameter
space, $C^\infty<1/2$, consistent with the bound of Proposition~3 for a single charger. At $|F|=1$, the maximum capacity increases with $N_A$, while the corresponding optimal coupling moving towards smaller $|J|$. The increments in maximum $C^{\infty}$ decrease with $N_A$, indicating a saturation of the maximum
capacity with the number of chargers, similar to the low-drive behavior (see
Proposition~5) in the case of a single-qubit charger.

The advantage due to the shared reservoir at $|F|=1$ is shown in
Fig.~\ref{fig:HD_star}(b). Note that $\delta C^\infty>0$ at weak coupling,
$\delta C^\infty<0$ at intermediate coupling ($0.5\lesssim|J|\lesssim2.3$), and a
weak advantage ($\delta C^\infty\lesssim0.02$) reappears at strong coupling
before $\delta C^\infty$ vanishes as $|J|\to\infty$, implying that the three regimes of Table~\ref{tab:dC_sign_NA}
persist beyond the perturbative regime.  The boundaries between the regimes, however, are shifted to much
larger couplings than in the low-drive limit, where Table~\ref{tab:dC_sign_NA}
places them at $|J|=\gamma_A/(2N_A)$ and $|J|=\gamma_B/2$ with $\gamma_A=\gamma_B=0.6$.  The shared reservoir is therefore advantageous over a wider range of couplings at strong drive and for more chargers, while the conditions in
Table~\ref{tab:dC_sign_NA} capture the correct qualitative structure but not the
positions of the boundaries beyond the low-drive limit. Note that the largest advantage occurs
as $|J|\to0$, where the battery is charged solely through the shared reservoir, while $C_0^\infty$ vanishes in its absence. The maximum $C^\infty$ at the low coupling grows with $N_A$.

\subsection{Single central charger qubit}
\label{subsec:multiqubit_battery}

We now explore charging multiple qubits with a weakly driven single-qubit central charger. Following Sec.~\ref{sec:multiqubit_charger}, we define $\boldsymbol{b}^{\pm}=\sum_{j\in B}b_j^{\pm}$, and drop $\mathcal{O}\left(|F|^{\ell>2}\right)$ terms (see Appendix~\ref{app:perturbation}) to obtain the following closed set of coupled linear equations: 
\begin{eqnarray}
    \langle \obj{b}^\pm\rangle_2^\infty&=&\frac{1}{\gamma_B}\bigg[2\text{Re}\left[R\langle a^+\obj{b}^-\rangle_2^\infty\right]-\Gamma\text{Re}\left[\langle \obj{b}^+\obj{b}^-\rangle_2^\infty\right]\bigg],\nonumber\\
    \langle \obj{b}^+\obj{b}^-\rangle_2^\infty&=&-\frac{2}{\chi_B}\text{Re}\left[S^*\langle a^+\obj{b}^-\rangle_2^\infty\right],\nonumber\\
    \langle a^+\obj{b}^-\rangle_2^\infty&=&\frac{2}{\xi_A+N_B\chi_B}\bigg(\text{i}|F|\langle \obj{b}^-\rangle_1^\infty-R^*\langle \obj{b}^+\obj{b}^-\rangle_2^\infty\nonumber\\&&-N_BS\langle a^+a^-\rangle_2^\infty\bigg),\nonumber\\
    \langle a^+a^-\rangle_2^\infty&=&\frac{2}{\xi_A}\left(\text{Re}\left[\text{i}|F|\langle a^-\rangle_1^\infty\right]-\text{Re}\left[R\langle a^+\obj{b}^-\rangle_2^\infty\right]\right),\nonumber\\
    \langle \obj{b}^-\rangle_1^\infty&=&-\frac{2S}{\chi_B}\langle a^-\rangle_1^\infty,\nonumber\\
    \langle a^-\rangle_1^\infty&=&-\frac{2}{\xi_A}\left[R\langle \obj{b}^-\rangle_1^\infty+\text{i}|F|\right].
\end{eqnarray}
where $R$ and $S$ are as defined in Eq.~(\ref{eq:RS}). Solving these leads to
\begin{eqnarray}
        C^\infty &=& \frac{4|F|^2\left[4|J|^2+\Gamma^2+4|J|\Gamma\sin\alpha\right]}{N_B^2\left[\left(\xi_A\chi_B+4|J|^2-\Gamma^2\right)^2+16|J|^2\Gamma^2\cos^2\alpha\right]},
        \label{eq:steady_state_energy_NB_battery}
\end{eqnarray}
where $E^{\max}=N_Bh$ is used. Clearly, for fixed $N_B$, $|J|$, $|F|$, $\Gamma$, and $\gamma_{A,B}$, $C^\infty$ is maximized at $\alpha=2n\pi+\pi/2$, $n\in\mathbb{Z}$, having a value 
\begin{eqnarray}
    C^{\infty}= \frac{4|F|^{2}u_+^{2}}{N_B^2\left(\xi_A\chi_B-u_+u_-\right)^{2}}.
    \label{eq:Cmax_2}
\end{eqnarray}
Also, at  $\alpha=2n\pi+\pi/2$ ($n\in\mathbb{Z}$) and for a fixed $N_B$, $C^\infty$ is maximum at $|J|_0=(\sqrt{\xi_A\chi_B}-\Gamma)/2$, with the maximum value given as a function of $N_B$ by 
\begin{eqnarray}
    C_{\max}^\infty(N_B)
    &=& \frac{|F|^2}{4N_B^2|J|_0^2}.
\end{eqnarray}  
Therefore, $C_{\max}^{\infty}\to 0$ as $\sim N_B^{-2}$ when $N_B\to\infty$. Further, proceeding as in Sec.~\ref{subsec:low_drive}, one can determine the condition for positivity of $\delta C^\infty$ as
\begin{eqnarray}
    \frac{u_+}{2|J|}>\frac{N_B(\xi_A\chi_B-u_+u_-)}{\gamma_A\gamma_B+4N_B|J|^2},
\end{eqnarray}
which leads to similar conditions on $|J|$ as given in Table~\ref{tab:dC_sign_NA}, but with $\gamma_A/N_A\to\gamma_A$ and $\gamma_B\to\gamma_B/N_B$.  It also establishes  $\min(\gamma_A,\gamma_B/N_B)\leq 2|J|_0\leq \max(\gamma_A,\gamma_B/N_B)$, leading to $\delta C^\infty<0$ at $|J|=|J|_0$.

These results can be summarized in the following proposition. 

\prop{6} \emph{For $N_B$ qubits as batteries and $1$ qubit as charger arranged in a star network with the charger connected to all battery qubits, with fixed $|F|\ll |J|$, $C^\infty$ and $\delta C^\infty$ exhibit the following features at $\alpha=2n\pi+\pi/2$ ($n\in\mathbb{Z}$).
\begin{enumerate}
    \item $C^{\infty}$ is maximized at $|J|_0=\left(\sqrt{\xi_A\chi_B}-\Gamma\right)/2$, but without any advantage over the collective-reservoir-free limit. 
    \item With $N_B\to\infty$,  the maximum $C^{\infty}$ decays as $\sim N_B^{-2}$.  
    \item With all other parameters fixed, and for $\Gamma>0$, $\delta C^{\infty}<0$ as long as $\min(\gamma_A,\gamma_B/N_B)< 2|J|< \max(\gamma_A,\gamma_B/N_B)$. 
\end{enumerate}}

\section{Conclusions and outlook}
\label{sec:conclusion}

In this paper, we investigated the charging of a many-body quantum battery in contact with an externally driven many-body charger, where each DoF in both the battery and the charger is coupled to a local dissipative bath, and the whole battery-charger system is immersed in a shared engineered reservoir. We quantified the performance of the battery via its steady-state capacity in terms of energy, ergotropy, and daemonic ergotropy, and introduced the bound capacity to account for the stored energy that cannot be extracted as work. Specifically,  we addressed two questions: (a) when the shared reservoir helps, and (b) whether complete nonreciprocity between the battery and the charger is necessary for optimal charging. 

For DoFs corresponding to qubits arranged on a lattice, we proved the steady state to be unique for arbitrary lattices, thereby making the steady-state performance independent of the initial state of the battery. For a single-qubit battery charged by a weakly driven single-qubit charger, using a perturbative approach in the external driving strength, we proved the Dzyaloshinskii-Moriya interaction between the battery and the charger to be optimal, and showed that the optimal coupling strength coincides with the nonreciprocal point only for a specific balance between the local and the collective dissipation rates, thereby demonstrating that complete nonreciprocity is, in general, not required for optimal charging. We further showed that the shared reservoir enhances the battery capacity over the shared-reservoir-free scenario when the coherent coupling is either weaker or stronger than both local dissipation rates, while it is detrimental at intermediate couplings, which include the capacity optimum. 

For the two-qubit battery-charger setup in the strong-drive limit, the battery capacity is bounded by $1/2$, while the shared reservoir is beneficial only at weak coupling, so that the maximum capacity over all values of $|J|$ and the advantage due to the shared reservoir cannot be achieved simultaneously in either of the low-, or the high-drive limit. Our analysis of the bound capacity further revealed (a) that the couplings maximizing the stored energy and the extractable work differ, (b) that the shared reservoir can reduce the fraction of the stored energy locked in the battery even when it reduces the stored energy itself, and (c) that measurements on the charger, via daemonic ergotropy, can unlock energy that is otherwise inaccessible. For qubits arranged in a star network, we showed that the maximum capacity of a single central battery qubit charged by $N_A$ weakly driven charger qubits saturates as $\sim N_A^{-1}$ in the presence of the shared reservoir, in contrast to the linear growth in its absence, while that of $N_B$ battery qubits charged by a single charger qubit decays as $\sim N_B^{-2}$. 

Our results open several directions for future investigation. Including intra-species interactions within the charger and the battery, as well as inhomogeneous drives, couplings, and dissipation rates, would allow one to probe the role of many-body effects, such as competing interactions and frustration, in the reciprocity landscape for DoFs arranged on a  complete bipartite graph. Moreover, despite initial steps in this direction~\cite{Liu2026arX}, whether complete nonreciprocity can be achieved on incomplete bipartite graphs, and how it affects the optimal performance of quantum batteries defined on them,
remains largely unexplored, the major difficulty being the requirement of the link-resolved engineered shared reservoir, as we point out in
Appendix~\ref{app:reciprocity}.  Extending the analysis beyond the steady state to the charging dynamics would clarify how the shared reservoir affects the charging power and time, while going beyond the low-drive perturbative regime analytically remains an open problem. Further, optimizing the measurement on the charger to minimize the bound capacity, and investigating non-Markovian or structured shared reservoirs, may provide additional routes to improve work extraction. Since Dzyaloshinskii-Moriya-type and nonreciprocal interactions between qubits can now be engineered in superconducting circuits~\cite{Ren2025,Zhang2026}, the setups considered here are within reach of current experimental platforms, providing a natural testbed for our predictions.

\acknowledgments

A.K.P thanks Prithvi Narayan P and Prasanna Venkatesh B for useful discussions, and acknowledges the support from the Anusandhan National Research Foundation (ANRF) of the Department of Science and Technology (DST), India, through the Core Research Grant (CRG) (File No. CRG/2023/001217, Sanction Date 16 May 2024). 


\appendix 

\section{Ergotropy and daemonic ergotropy}
\label{app:ergotropy_definitions}

Let the state of a $D$-dimensional quantum system at an arbitrary time $t$ during the time-evolution be given by $\rho$. The ergotropy, $\mathcal{E}$, of $\rho$ w.r.t. the system Hamiltonian $H$ quantifies the maximum extractable work from $\rho$ w.r.t. $H$ under cyclic unitaries, and is defined as
\begin{eqnarray}
    \mathcal{E}&=&\mathcal{E}(\rho(t))\nonumber\\&=&\text{Tr}[\rho H]-\min_{\{V\}}\text{Tr}\left[V\rho V^\dagger H\right], 
\end{eqnarray}
where the minimization runs over the complete set of cyclic unitary matrices $\{V\}$ on the Hilbert space of the system. It can be calculated as
\begin{eqnarray}
    \mathcal{E}&=&\text{Tr}\left[H\left(\rho-\tilde{\rho}\right)\right],  
\end{eqnarray}
where
\begin{eqnarray}
    \tilde{\rho}=\sum_{\beta=0}^{D-1}q_\beta\ket{\beta}\bra{\beta},
\end{eqnarray}
is a passive state obtained from $\rho$, with $\{q_{\beta}\}$ being the eigenvalues of $\rho$ arranged in descending order (i.e., $q_{\beta}\geq q_{\beta+1}$, $0\leq \beta\leq D-1$), and $\{\ket{\beta}\}$ being the energy eigenvectors of $H$ having energy eigenvalues $E_\beta$ such that $E_\beta\leq E_{\beta+1}$, $0\leq \beta\leq D-1$. Note that $\rho$, and subsequently $\mathcal{E}$, are functions of $t$.   

We now consider a bipartite scenario where $\rho$ denotes the state of a battery-charger setup (see Sec.~\ref{subsec:system}). We denote the battery state by $\rho_B=\text{Tr}_A\rho$, which can be written as 
\begin{eqnarray}
    \rho_B=\sum_{k}p^k\rho_B^k,
\end{eqnarray}
for each member of an infinite set of ensembles $\left\{\rho_B^k=\text{Tr}_A\left[\rho^k\right],p^k\right\}$, and has an ergotropy $\mathcal{E}$. A member in this set is generated by performing a projection measurement $M_A$ on the subsystem $A$ in the state $\rho$, defined by the projection operators $\{P_A^{k}=\ket{k}\bra{k}\}$ corresponding to the measurement outcomes $k$, leading to 
\begin{eqnarray}
    \rho\to\rho^k=\left(P_A^k\rho P_A^k\right)/p^k,
\end{eqnarray}
with 
\begin{eqnarray}
    p^k=\text{Tr}\left[P_A^k\rho P_A^k\right].
\end{eqnarray}
At an arbitrary time $t$, the daemonic ergotropy corresponding to each measurement $M_A$  is defined as
\begin{eqnarray}
    \overline{\mathcal{E}}(M_A)=\sum_{k}p^{k} \mathcal{E}^{k}(M_A),
\end{eqnarray}
with $\mathcal{E}^{k}(M_A)$ being the ergotropy of  $\rho_B^{k}$, where $\{M_A\}$ is the complete set of possible projection measurements on $A$. We will be interested in the maximum daemonic ergotropy
\begin{eqnarray}
    \overline{\mathcal{E}}_0&=&\max_{\{M_A\}}\overline{\mathcal{E}}(M_A),
\end{eqnarray}
maximized over the complete $\{M_A\}$, while the convexity of ergotropy leads to 
\begin{eqnarray}
    \overline{\mathcal{E}}_0\geq \overline{\mathcal{E}}(M_A)\geq \mathcal{E}\;\forall\; M_A.
\end{eqnarray}

\section{Existence of unique steady state}
\label{app:steady_state_existence}

For a finite dimensional Hilbert space, each super-operator $\mathcal{L}(.)$ corresponds to at least one stationary state~\cite{Baumgartner2008a,Baumgartner2008b} for which $\tilde{\rho}(t=0)=\tilde{\rho}(t=\infty)$. Further, the steady states, $\tilde{\rho}(t=\infty)$, of
\emph{irreducible} systems are unique and faithful~\cite{Zhang2024}. Given the set $\{L_i\}$ of Lindblad operators  corresponding to $\mathcal{L}(.)$, and defining 
\begin{eqnarray}
    K=\text{i}H_S+\frac{1}{2}\sum_i L_i^\dagger L_i,
\end{eqnarray}
with $H_S$ being the system Hamiltonian,  a sufficient condition to test irreducibility is to find a subset $\mathbf{G}$ of the algebra $\mathbf{A}$ generated by $\{L_i,K\}$ having a self-adjoint span, and a trivial commutant~\cite{Zhang2024}. Note that the subset $\mathbf{G}$ has a self-adjoint span if an operator $\mathcal{O}\in\mathbf{G}$ implies $\mathcal{O}^\dagger\in\mathbf{G}$. On the other hand, for $\mathbf{O}(\mathbf{H})$ being the set of all linear operators defined on the system Hilbert space $\mathbf{H}$,  the commutant $\mathbf{X}^\prime\subseteq\mathbf{O}(\mathbf{H})$ of a set $\mathbf{X}\subseteq \mathbf{O}(\mathbf{H})$ is the set of operators $\mathbf{Y}\subseteq \mathbf{O}(\mathbf{H})$ such that $[X,Y]=0$ $\forall$ $X\in \mathbf{X}$ if and only if $Y\in \mathbf{Y}$.

We now prove the irreducibility of the battery-charger setup considered in this paper, where each DoF corresponds to a qubit, assuming Eq.~(\ref{eq:irreducibility_validity}) to hold, along with $|F_i|\neq 0\forall i\in A$ (see Appendix~\ref{app:perturbation} for the case of $|F_i|=0\forall i\in A$). 
Let us consider   
\begin{eqnarray}
L_i&=&\sqrt{\gamma_i}a_i^-,\nonumber\\
L_j&=&\sqrt{\gamma_j}b_j^-, \nonumber\\
L_{AB}&=&\sqrt{\Gamma}\left(\sum_{i\in A}\mathcal{A}_ia_i^-+\sum_{j\in B}\mathcal{B}_jb_j^-\right),
\end{eqnarray}
$\forall i\in A,\forall j\in B$,  leading to 
\begin{widetext}
\begin{eqnarray}    
K&=&\text{i}\sum_{\langle i,i^\prime\rangle\in G}\left(g_{i,i^\prime}a_i^+a_{i^\prime}^-+g^*_{i,i^\prime}a_i^-a_{i^\prime}^+\right)+\text{i}\sum_{\langle j,j^\prime\rangle\in G}\left(g_{j,j^\prime}b_j^+b_{j^\prime}^-+g^*_{j,j^\prime}b_j^-b_{j^\prime}^+\right)+\text{i}\sum_{\langle i,j\rangle\in G}\left(J_{i,j}a_i^+b_j^-+J^*_{i,j}a_i^-b_j^+\right)\nonumber\\&&+\sum_{i\in A}\frac{\gamma_i}{2}a_i^+a_i^-+\sum_{j\in B}\frac{\gamma_j}{2}b_j^+b_j^-+\text{i}\sum_{i \in A}\left(|F_{i}|a_i^++|F_{i}|a_i^-\right)+\frac{\Gamma}{2}\bigg[\sum_{i,i^\prime\in A}a_i^+a_{i^\prime}^-+\sum_{j,j^\prime\in B}b_j^+b_{j^\prime}^-+\sum_{i\in A}\sum_{j\in B}\left(a_i^+b_{j}^-+a_i^-b_{j}^+\right)\bigg].
\end{eqnarray}    
\end{widetext}
Noting that $a_i^-=L_i/\sqrt{\gamma_i}$ and $b_j^-=L_j/\sqrt{\gamma_j}$, and identifying
\begin{eqnarray}
    \mathbf{G}=\{a_i^\pm,b_j^\pm\};\quad \forall i\in A, j\in B,
\end{eqnarray}
one can show that $I\in \mathbf{A}$, since  
\begin{eqnarray}
    I&=&M^2-2\text{i}M,
\end{eqnarray}
where $M=|F_i|^{-1}\left[a_i^-\mathcal{O}_1+\text{i}|F_i|\right]=\{K,a_i^-\}-\mathcal{O}_0\in \mathbf{A}$, with 
\begin{eqnarray}
\mathcal{O}_0&=&\sum_{i^\prime\in N(i)}g_{i,i^\prime}a_{i^\prime}^--\sum_{i^\prime\in N(i)}g^*_{i,i^\prime}a_{i^\prime}^-\nonumber\\&&+\sum_{j \in N(i)}J_{i,j}b_{j}^-+2\text{i}\sum_{\underset{i\ne i^\prime}{i^\prime \in A}}|F_{i^\prime}|a_i^-a_{i^\prime}^-\nonumber\\&&+\frac{\gamma_i}{2}a_i^-+\frac{\Gamma}{2}\bigg[\sum_{\underset{i\ne i^\prime}{i^\prime \in A}}a^-_{i^\prime}+a_i^-+\sum_{j\in B}b_j^-\bigg],
\end{eqnarray}
and
\begin{widetext}
\begin{eqnarray}
    \mathcal{O}_1 &=& 2\text{i}\Bigg[\sum_{\underset{i^\prime,i^{\prime\prime}\ne i}{\langle i^\prime,i^{\prime\prime}\rangle\in G}}g_{i^\prime,i^{\prime\prime}}a_{i^\prime}^+a_{i^{\prime\prime}}^-+\sum_{\underset{i^\prime,i^{\prime\prime}\ne i}{\langle i^\prime,i^{\prime\prime}\rangle\in G}}g^*_{i^\prime,i^{\prime\prime}}a_{i^\prime}^-a_{i^{\prime\prime}}^++\sum_{\langle j,j^\prime\rangle \in G}\left(g_{j,j^\prime }b_j^+b^-_{j^\prime}+g^*_{j,j^\prime }b_j^-b^+_{j^\prime}\right)+\sum_{\underset{i^\prime\ne i}{\langle i^\prime,j\rangle \in G}}J_{i^\prime,j} a^+_{i^\prime}b_j^-\nonumber\\
    &&+\sum_{\underset{i^\prime\ne i}{\langle i^\prime,j\rangle \in G}}J^*_{i^\prime,j} a^-_{i^\prime}b_j^++\sum_{\underset{i^\prime\ne i}{i^\prime \in A}}|F_{i^\prime}|a^+_{i^\prime}\Bigg]+\sum_{\underset{i^\prime \ne i}{i^\prime \in A}}\gamma_{i^\prime}a_{i^\prime}^+a_{i^\prime}^-+\sum_{j\in B}\gamma_jb_j^+b_j^-+\Gamma\Bigg[\sum_{j,j^\prime \in B} b_j^+b_{j^\prime}^-+\sum_{\underset{i^\prime,i^{\prime\prime}\ne i}{i^\prime,i^{\prime\prime}} \in A}a_{i^\prime}^+a_{i^{\prime\prime}}^-\nonumber\\
    &&+\sum_{\underset{i^\prime\ne i}{i^\prime\in A}}\sum_{j\in B} a_{i^\prime}^+b_j^-+\sum_{\underset{i^\prime\ne i}{i^\prime\in A}}\sum_{j\in B} a_{i^\prime}^-b_j^+\Bigg],
\end{eqnarray}
\end{widetext}
leading further to $M|F|-\text{i}|F|=a_i^-\mathcal{O}_1\in \mathbf{A}$. Further, 
\begin{eqnarray}
    &&(\text{i}|F|)^{-1}\bigg(a_i^-K-2a^-_{i}\mathcal{O}_1-\text{i}\sum_{\underset{i^\prime\ne i}{i^\prime\in A}}|F_{i^\prime}|a_{i^\prime}^-a_i^--\frac{\gamma_i}{2}a_i^-+\frac{\Gamma}{2}a_i^-\bigg)\nonumber\\&=&\left[(\text{i}|F|)^{-1}\mathcal{O}_2+I\right]a_i^-a_i^+\nonumber\\&\in&\mathbf{A},
\end{eqnarray}
with $\mathcal{O}_2$ given by
\begin{eqnarray}
    \mathcal{O}_2&=&\text{i}\bigg[\sum_{i^\prime\in N(i)}g_{i,i^\prime}a_{i^\prime}^-+\text{i}\sum_{i^\prime\in N(i)}g^*_{i,i^\prime}a_{i^\prime}^-+\text{i}\sum_{j\in N(i)}J_{i,j}b_j^-\bigg]\nonumber\\&&+\frac{\Gamma}{2}\bigg[\sum_{i^\prime\in N(i)}a^-_{i^\prime}+\sum_{j\in B}b_j^-\bigg],
\end{eqnarray}
where we have denoted the neighborhood of the DoF $i$ by $N(i)$. Since $\mathcal{O}_2$ is nilpotent (i.e., $\mathcal{O}_2^N=0, N\in \mathbf{Z^+}$), 
\begin{eqnarray}
    a_i^-a_i^+ 
    &=&\bigg[\sum_{n=0}^{N-1}\left\{(\text{i}|F|)^{-1}\mathcal{O}_2\right\}^n\bigg](\text{i}|F|)^{-1}\bigg[a_i^-K-2a^-_{i}\mathcal{O}_1\nonumber\\&&\quad-\text{i}\sum_{\underset{i^\prime\ne i}{i^\prime\in A}}|F_{i^\prime}|a_{i^\prime}^-a_i^--\frac{\gamma_i}{2}a_i^-+\frac{\Gamma}{2}a_i^-\bigg].
\end{eqnarray}
Also, 
\begin{eqnarray}
    a_i^+&=&(\text{i}|F|)^{-1}\left[I+(\text{i}|F|)^{-1}\mathcal{O}_3\right]^{-1}Ka_i^-a_i^+\nonumber\\&\in& \mathbf{A},
\end{eqnarray}
since
\begin{eqnarray}
    \mathcal{O}_3 &=& \frac{\mathcal{O}_1a_i^-}{2}+\text{i}\sum_{i^\prime\in N(i)}g_{i^\prime,i}^*a_{i^\prime}^-+\text{i}\sum_{j\in N(i)}J_{i,j}b_j^-\nonumber\\&&+\sum_{i^\prime\in N(i)}g_{i,i^\prime}a_{i^\prime}^-+\text{i}\sum_{i^\prime\in A}|F_{i^\prime}|a_{i^\prime}^-a_i^-\nonumber\\&&+\frac{\Gamma}{2}\bigg[\sum_{\underset{i^\prime\ne i}{i^\prime\in A}}a_{i^\prime}^-+\sum_{j\in B}b_j^-\bigg]\nonumber\\&&\in \mathbf{A}
\end{eqnarray}
is also nilpotent, and 
\begin{eqnarray}
\left[I+(\text{i}|F|)^{-1}\mathcal{O}_3\right]^{-1}&=&\sum_{n=0}^{N-1}(\text{i}|F|)^{-n}\mathcal{O}_3^n\nonumber\\&\in& \mathbf{A}.    
\end{eqnarray}
One can further define 
\begin{eqnarray}
    b_j^z = \left(\text{i}J^*_{i,j}+\frac{\Gamma}{2}\right)^{-1}a_i^z\left[[a_i^+,\mathcal{O}_4],b_j^-\right],
\end{eqnarray}
with
\begin{eqnarray}
    \mathcal{O}_4&=&K-\sum_{i\in A}\frac{\gamma_i}{2}a_i^+a_i^-\nonumber\\&&-\text{i}\Bigg[\sum_{\langle i,i^\prime\rangle\in G}\left(g_{i,i^\prime}a_i^+a_{i^\prime}^-+g^*_{i,i^\prime}a_i^-a_{i^\prime}^+\right)\nonumber\\&&+\sum_{\langle i,j\rangle\in G}J_{i,j}a_i^+b_j^-+\sum_{i \in A}\left(|F_{i}|a_i^++|F_{i}|a_i^-\right)\Bigg]\nonumber\\&&-\frac{\Gamma}{2}\left[\sum_{i,i^\prime\in A}a_i^+a_{i^\prime}^--\sum_{i\in A}\sum_{j\in B}a_i^+b_{j}^-\right],
\end{eqnarray}
and subsequently 
\begin{eqnarray}
    b_j^+b_j^-=\frac{b_j^z+I}{2},
\end{eqnarray}
and
\begin{eqnarray}
    b_j^+=-\left(\text{i}J^*_{i,j}+\frac{\Gamma}{2}\right)^{-1}\left\{a_i^+\left[\left\{\mathcal{O}_4,a_i^+\right\},b_j^+b_j^-\right],a_i^-\right\}.
\end{eqnarray}
Hence, each element of $\mathbf{G}$ can be constructed from $\mathbf{A}$.

Next, let us take a general operator
\begin{eqnarray}
    X &=& \bigotimes_{k=1}^{N}X_k,
\end{eqnarray}
with 
\begin{eqnarray}
    X_k=\sum_{\alpha_k,\beta_k =0}^{1}\ket{\alpha_k}\bra{\beta_k}; k = 1,...,N-1,
\end{eqnarray}
and 
\begin{eqnarray}
    X_N = \sum_{\alpha_N,\beta_N =0}^{1} C_{\alpha_1,\beta_1,...,\alpha_N,\beta_N}\ket{\alpha_N}\bra{\beta_N},
\end{eqnarray}
where $N=N_A+N_B$ is the total number of qubits in the system.  Assuming $X$ to be an element of the commutant of $\mathbf{G}$, 
\begin{eqnarray}
    [X,b_N^\pm] &=& X_1\otimes...\otimes X_{N-1}\otimes\left[ X_N ,\sigma_N^\pm\right]=0.
\end{eqnarray}
In this case, we can see that $[X_N,\sigma_N^\pm]=0$ and $[X_N,\sigma_N^z]=0$. Since, $X_N$ commutes with each element of the set $\mathbf{S} = \{\sigma^\pm,\sigma^z\}$ representing the full spin-$\frac{1}{2}$ algebra corresponding to the Hilbert space of the $N$th qubit and since the spin-$\frac{1}{2}$ representation is irreducible, by Schur's lemma~\cite{Schur1905}, $X_N=\lambda_{\alpha_1,\beta_1,...,\alpha_{N-1},\beta_{N-1}} I_N$, $\lambda_{\alpha_1,\beta_1,...,\alpha_{N-1},\beta_{N-1}}\in \mathbb{C}$, leading to $X=X_1\otimes X_2\otimes...\otimes X_{N-1}\otimes I_N$, with 
\begin{eqnarray}
    X_{N-1} = \sum_{\alpha_{N-1},\beta_{N-1} =0}^{1}\lambda_{\alpha_1,\beta_1,...,\alpha_{N-1},\beta_{N-1}}\ket{\alpha_{N-1}}\bra{\beta_{N-1}},
\end{eqnarray}
Repeating this till the $N_A^{th}$ qubit, followed by taking the commutation with $a_{N_A}^\pm,a^z_{N_A}$ and continuing this to all the qubits in subsystem A as well to finally get $X=\lambda I$, where $I = I_1\otimes I_2\otimes...\otimes I_N$ and thereby proving that $\mathbf{G}$ has a trivial commutant. Also, the self-adjoint property of $\mathbf{G}$ is apparent since  $\mathcal{O}\in \mathbf{G}$ implies  $\mathcal{O}^\dagger\in \mathbf{G}$.

\subsection{Irreducibility of a driven-dissipative qubit}
\label{app:irreducibility_single_qubit}

Let us separately consider a qubit connected to a dissipative bath with dissipation rate $\gamma$, and driven by a field of strength $F$ (see Sec.~\ref{sec:charging_a_qubit}), for which the jump operator is $L=\sqrt{\gamma}a^-$, and $K=\text{i}|F|a^++\text{i}|F|a^-+\frac{\gamma}{2}a^+a^-$. We consider the set $\mathbf{G}=\{a^\pm\}$, and derive 
\begin{eqnarray}
    a^-&=&\frac{L}{\sqrt{\gamma}},\nonumber\\
    a^+a^-&=&-\text{i}|F|^{-1}Ka^-,\nonumber\\
    a^+&=&(\text{i}|F|)^{-1}(K-\text{i}|F|a^-)(I+\text{i}(2|F|)^{-1}\gamma a^-),
\end{eqnarray}
and 
\begin{eqnarray}
        I&=& (\text{i}|F|)^{-1}(K-\text{i}|F|a^--\frac{\gamma}{2}a^+a^-),
\end{eqnarray}
from the algebra of $K$ and $L$.
Next, we consider an arbitrary single qubit operator $X=\sum_{\alpha,\beta}C_{\alpha,\beta}\ket{\alpha}\bra{\beta}$, which, being an element of the commutant of $\mathbf{G}$, commutes with $a^\pm$. Since $X$ commutes with the complete spin-$\frac{1}{2}$ algebra $\{a^\pm,a^z\}$, and since the spin-$\frac{1}{2}$ representation is irreducible, by Schur's lemma $X=\lambda I, \lambda \in \mathbb{C}$, demonstrating that $\mathbf{G}$ has a trivial commutant.  Further, $\mathbf{G}$ is self adjoint by definition, thereby ensuring the existence of a unique steady state.

\begin{figure}
    \centering
    \includegraphics[width=\linewidth]{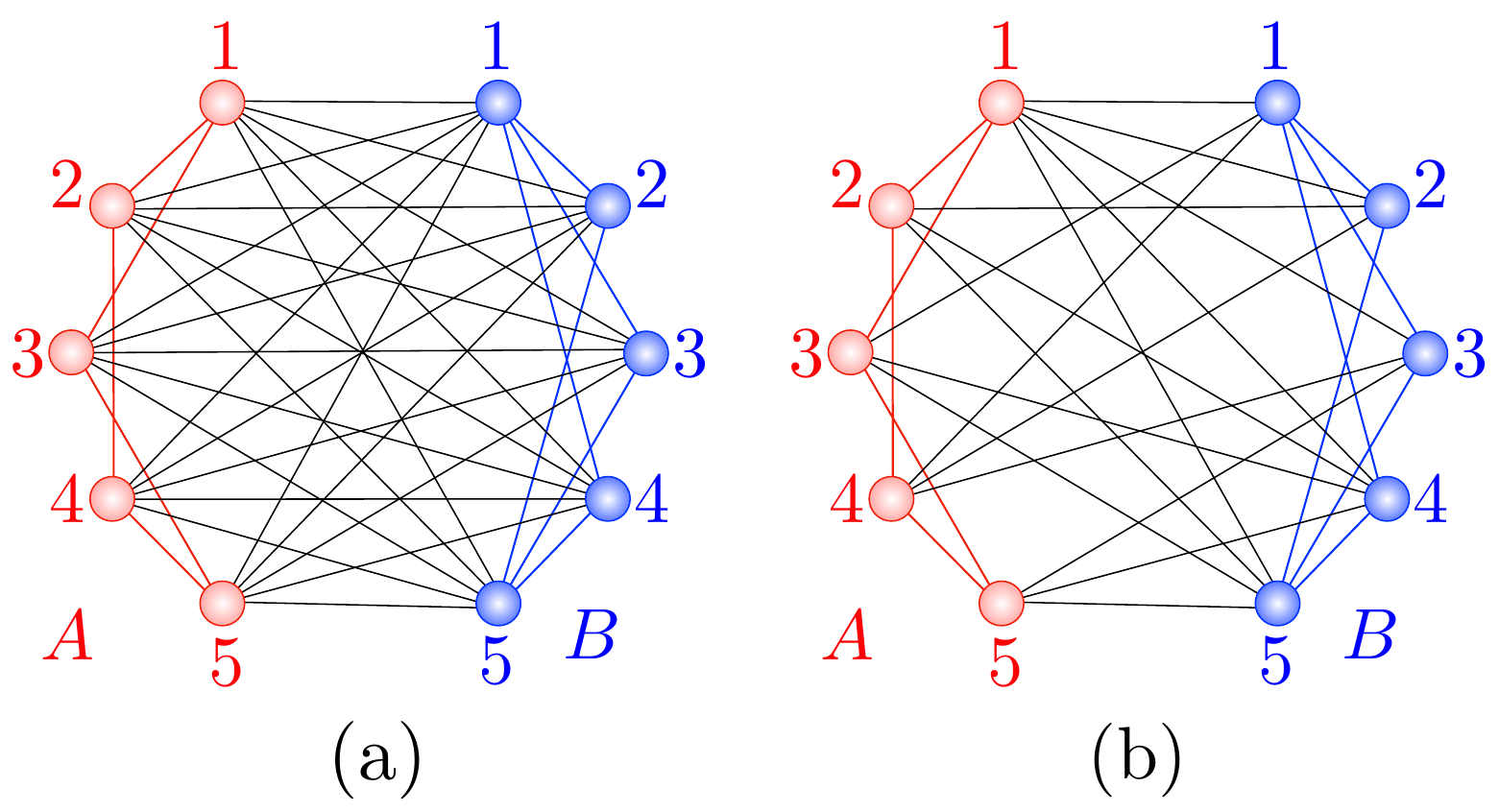}
    \caption{Examples of (a) a complete, and (b) an incomplete bipartite graph.}
    \label{fig:graphs}
\end{figure}

\section{The reciprocity landmarks} 
\label{app:reciprocity}

Assuming the observable to have non-trivial support only on $A$ and denoting it with $\mathcal{O}_A$, from Eq.~(\ref{eq:Opeator_evolution_gen_1}), one obtains
\begin{widetext}
\begin{eqnarray}
\label{eq:Opeator_evolution_A_final}
\frac{d\left\langle\mathcal{O}_A\right\rangle}{dt}
&=&\text{i}\left\langle[\tilde{H}_0+\tilde{H}_A,\mathcal{O}_A]\right\rangle+\sum_{i\in A}\frac{\Gamma |\mathcal{A}_i|^2+\gamma_i}{2} \left(\left\langle a_i^+[\mathcal{O}_A,a_i^-]\right\rangle+\left\langle [a_i^+,\mathcal{O}_A]a_i^-\right\rangle\right)\nonumber\\&+&\frac{\Gamma}{2}\sum_{\underset{i\ne i^\prime}{i,i^\prime \in A}}\mathcal{A}_i^*\mathcal{A}_{i^\prime}\left[\left\langle a_i^+[\mathcal{O}_A,a_{i^\prime}^-]\right\rangle+\left\langle [a_i^+,\mathcal{O}_A]a_{i^\prime}^-\right\rangle\right]+\sum_{\langle i,j\rangle\in G}\left[R_{i,j}\left\langle[a_i^+,\mathcal{O}_A]b_j^-\right\rangle-R_{i,j}^*\left\langle[a_i^-,\mathcal{O}_A]b_j^+\right\rangle\right]\nonumber\\&+&\frac{\Gamma}{2}\sum_{\langle i,j\rangle\not{\in} G}|\mathcal{C}_{i,j}|\text{e}^{\text{i}\phi_{i,j}}\left[\left\langle[a_i^+,\mathcal{O}_A]b_j^-\right\rangle-\text{e}^{-2\text{i}\phi_{i,j}}\left\langle[a_i^-,\mathcal{O}_A]b_j^+\right\rangle\right],
\end{eqnarray}
\end{widetext}
with 
\begin{eqnarray}
    R_{i,j}&=&\text{i}|J_{i,j}|\text{e}^{\text{i}\theta_{i,j}}+\frac{\Gamma |\mathcal{C}_{i,j}|\text{e}^{\text{i}\phi_{i,j}}}{2}\;\;\forall\langle i,j\rangle \in G,
\end{eqnarray}
where we have defined  $\mathcal{A}_i^*\mathcal{B}_j=|\mathcal{C}_{i,j}|\text{e}^{\text{i}\phi_{i,j}}$. In a similar fashion, for an observable $\mathcal{O}_B$ on $B$, one obtains
\begin{widetext}
\begin{eqnarray}
\label{eq:Opeator_evolution_B_final}
\frac{d\left\langle\mathcal{O}_B\right\rangle}{dt}
&=&\text{i}\left\langle[\tilde{H}_0+\tilde{H}_B,\mathcal{O}_B]\right\rangle+\sum_{j\in B}\frac{\Gamma |\mathcal{B}_j|^2+\gamma_j}{2} \left(\left\langle b_j^+[\mathcal{O}_B,b_j^-]\right\rangle+\left\langle [b_j^+,\mathcal{O}_B]b_j^-\right\rangle\right)\nonumber\\&+&\frac{\Gamma}{2}\sum_{\underset{j\ne j^\prime}{j,j^\prime \in A}}\mathcal{B}_j^*\mathcal{B}_{j^\prime}\left[\left\langle b_j^+[\mathcal{O}_B,b_{j^\prime}^-]\right\rangle+\left\langle [b_j^+,\mathcal{O}_B]b_{j^\prime}^-\right\rangle\right]+\sum_{\langle i,j\rangle\in G}\left[S_{i,j}\left\langle a_i^-[b_j^+,\mathcal{O}_B]\right\rangle-S_{i,j}^*\left\langle a_i^+[b_j^-,\mathcal{O}_B]\right\rangle\right]\nonumber\\&+&\frac{\Gamma}{2}\sum_{\langle i,j\rangle\not{\in} G}|\mathcal{C}_{i,j}|\text{e}^{-\text{i}\phi_{i,j}}\left[\left\langle a_i^-[b_j^+,\mathcal{O}_B]\right\rangle-\text{e}^{2\text{i}\phi_{i,j}}\left\langle a_i^+[b_j^-,\mathcal{O}_B]\right\rangle\right],
\end{eqnarray}
\end{widetext}
with 
\begin{eqnarray}
S_{i,j}&=&\text{i}|J_{i,j}|\text{e}^{-\text{i}\theta_{i,j}}+\frac{\Gamma |\mathcal{C}_{i,j}|\text{e}^{-\text{i}\phi_{i,j}}}{2}\;\;\forall \langle i,j\rangle\in G,
\end{eqnarray}
using, by definition, $\mathcal{A}_i\mathcal{B}_j^*=|\mathcal{C}_{i,j}|\text{e}^{-\text{i}\phi_{i,j}}$. 

\subsection{Reciprocity in complete bipartite graph}
\label{app:complete_bipartite}

We first consider a complete bipartite graph depicting the connectivity between the DoFs. In such a graph, all nodes can be divided into two species $A$ and $B$, and the inter-species connectivity is such that each node in $A$ is connected to each node in $B$ (see Fig.~\ref{fig:graphs} (a)), irrespective of the intra-species connectivity. Note that for such graphs, the last terms in both Eqs.~(\ref{eq:Opeator_evolution_A_final}) and (\ref{eq:Opeator_evolution_B_final}) vanish even for $\Gamma\neq 0$.  In Eq.~(\ref{eq:Opeator_evolution_A_final}), we demand $\mathcal{O}_A^\prime=[H_0,\mathcal{O}_A]$ to either be $0$, or have support only on $A$, thereby imposing important constraints on the possible forms of $H_0$. With this, a breaking of \emph{reciprocity} between $A$ and $B$ via a complete removal of the effect of subsystem $B$ on  $\langle\mathcal{O}_A\rangle$ takes place iff $R_{i,j} = 0\;\forall\; (i,j)$ pair, leading to the \emph{complete non-reciprocity} (CNR) condition for the complete bipartite graphs:
\begin{eqnarray}
    \alpha_{i,j}&=&2n\pi+\frac{\pi}{2},\nonumber\\
    |J_{i,j}|&=&\Gamma |\mathcal{C}_{i,j}|/2,
\end{eqnarray}
$\forall \langle i,j\rangle\in G$, with $n\in\mathbb{Z}$, where we have defined 
\begin{eqnarray}
    \alpha_{i,j}=\theta_{i,j}-\phi_{i,j}.
\end{eqnarray}
We refer to this as $\text{CNR}_{A\not\leftarrow B}$ (see Fig.~\ref{fig:parameter_space} for the simplest case of two-qubit graph).

Similarly, demanding $\mathcal{O}_B^\prime=[H_0,\mathcal{O}_B]$ in Eq.~(\ref{eq:Opeator_evolution_B_final}) to either be $0$, or have support only on $B$, a complete removal of the effect of $A$ on $\langle\mathcal{O}_B\rangle$ occurs when $S_{i,j}=0$ $\forall$ $(i,j)$ pair, leading to the  $\text{NR}_{A\not\to B}$ condition
\begin{eqnarray}
    \alpha_{i,j}&=&2n\pi+\frac{3\pi}{2},\nonumber\\
    |J_{i,j}|&=&\frac{\Gamma |\mathcal{C}_{i,j}|}{2},
\end{eqnarray}
$\forall \langle i,j\rangle\in G$, with $n\in\mathbb{Z}$ (see Fig.~\ref{fig:parameter_space} for an illustration in the case of the two-qubit graph). On the other hand, a \emph{complete reciprocity} (CR) between $A$ and $B$ is obtained when $R_{i,j}=S_{i,j}$, leading to the $\text{CR}_{A\leftrightarrow B}$ condition 
\begin{eqnarray}
    \theta_{i,j}&=&n\pi+(-1)^n\phi_{i,j},\nonumber\\ |J_{i,j}|&=&\frac{\Gamma|\mathcal{C}_{i,j}|}{2},
\end{eqnarray}
$\forall \langle i,j\rangle\in G$, with $n\in\mathbb{Z}$. 

In this paper, we focus only on complete bipartite graphs. In view of the CNR conditions corresponding to a complete bipartite graph, we are specifically interested in the case of $n$ being even ($n=2m,m\in\mathbb{Z}$),  leading to
\begin{eqnarray}
    \alpha_{i,j}&=&2m\pi,\nonumber\\ |J_{i,j}|&=&\frac{\Gamma|\mathcal{C}_{i,j}|}{2},
\end{eqnarray}
$\forall \langle i,j\rangle\in G$ (see Fig.~\ref{fig:parameter_space}). With a varying $|J_{i,j}|$ (i.e., varying $\Gamma$, or $|\mathcal{C}_{i,j}|$, or both), the CNR and CR points transform to lines, while further introduction of varying parameters leads to higher-dimensional CNR and CR manifolds.

\subsection{Reciprocity in an incomplete bipartite graph}
\label{app:incomplete_bipartite}

Note that the last terms in Eqs.~(\ref{eq:Opeator_evolution_A_final}) and (\ref{eq:Opeator_evolution_B_final}) can respectively be written as 
\begin{eqnarray}
    &&\sum_{\langle i,j\rangle\notin G}\left[R_{i,j}^{(0)}\left\langle[a_i^+,\mathcal{O}_A]b_j^-\right\rangle-R_{i,j}^{(0)*}\left\langle[a_i^-,\mathcal{O}_A]b_j^+\right\rangle\right],\nonumber
\end{eqnarray}
and 
\begin{eqnarray}
    &&\sum_{\langle i,j\rangle\notin G}\left[S_{i,j}^{(0)}\left\langle a_i^-[b_j^+,\mathcal{O}_B]\right\rangle-S_{i,j}^{(0)*}\left\langle a_i^+[b_j^-,\mathcal{O}_B]\right\rangle\right], \nonumber
\end{eqnarray}
where $R_{i,j}^{(0)}$ and $S_{i,j}^{(0)}$ are respectively $R_{i,j}$ and $S_{i,j}$ with $|J_{i,j}|=0$. For incomplete bipartite graphs (see Fig.~\ref{fig:graphs} (b)), the these terms do not vanish unless $\Gamma|C_{i,j}|=0$ for all
$\langle i,j\rangle\notin G$. For $\Gamma>0$ and $\mathcal{A}_i,\mathcal{B}_j\neq0$, complete nonreciprocity can only be reached for complete bipartite graphs with $|J_{i,j}|=\Gamma|C_{i,j}|/2$, and the same holds for CNR$_{A\not\to B}$. Physically, each missing coherent link leaves a purely dissipative channel with $|R_{i,j}|=|S_{i,j}|$ allowing the  battery to acts back on the charger with the same strength as the charger acts on
the battery. Since $C_{i,j}=\mathcal{A}_i^*\mathcal{B}_j$, this reservoir channel
cannot be suppressed on selected links by tuning a single collective jump
operator, subsequently making it impossible to achieve complete nonreciprocity unless the engineered reservoir is link-resolved. On the other hand, complete reciprocity remains attainable in the case of the incomplete bipartite graph via ensuring $R_{i,j}=S_{i,j}$, and $R_{i,j}^{(0)}=S_{i,j}^{(0)}$.

\section{Perturbative treatment of the quantum master equation}
\label{app:perturbation}

Consider a situation where the dynamics of a system described by the Hamiltonian $H_0$ is dictated by the quantum master equation 
\begin{eqnarray}
    \frac{d\rho}{dt}=\mathcal{L}_0(\rho)=-\text{i}\left[H_0,\rho\right]+\mathcal{D}(\rho),
    \label{eq:unperturbed_qme}
\end{eqnarray}
with $\mathcal{D}(.)$ representing the dissipators. Introduction of a perturbative component $H^\prime$ of strength $\lambda$ to the system Hamiltonian modifies   Eq.~(\ref{eq:unperturbed_qme}) to 
\begin{eqnarray}
    \frac{d\rho}{dt}=\mathcal{L}_0(\rho)+\lambda\mathcal{L}^\prime(\rho),
    \label{eq:perturbed_qme}
\end{eqnarray}
where $\mathcal{L}^\prime(\rho)=-\text{i}\left[H^\prime,\rho\right]$. This leads to 
\begin{eqnarray}
    \frac{d\langle \mathcal{O}\rangle}{dt}=\langle \mathcal{L}_0^\dagger(\mathcal{O})\rangle+\lambda\langle\mathcal{L}^{\prime\dagger}(\mathcal{O})\rangle,
    \label{eq:perturbed_operator_eqn}
\end{eqnarray}
for an arbitrary operator $\mathcal{O}$. Expanding in powers of $\lambda$,
\begin{eqnarray}        \langle\mathcal{O}\rangle&=&\sum_{\ell=0,1,2,\cdots}\lambda^\ell\langle\mathcal{O}\rangle_\ell,\nonumber\\
\langle\mathcal{L}^\dagger(\mathcal{O})\rangle&=&\sum_{\ell=0,1,2,\cdots}\lambda^\ell\langle\mathcal{L}^\dagger(\mathcal{O})\rangle_\ell,\nonumber\\ \langle\mathcal{L}^{\prime\dagger}(\mathcal{O})\rangle&=&\sum_{\ell=0,1,2,\cdots}\lambda^\ell\langle\mathcal{L}^{\prime\dagger}(\mathcal{O})\rangle_\ell,
\end{eqnarray}
and substituting in Eqs.~(\ref{eq:perturbed_qme}) and (\ref{eq:perturbed_operator_eqn}), one obtains 
\begin{eqnarray}        \frac{d\langle\mathcal{O}\rangle_0}{dt}&=&\langle\mathcal{L}_0^\dagger(\mathcal{O})\rangle_0,\nonumber\\
\frac{d\langle\mathcal{O}\rangle_\ell}{dt}&=& \langle\mathcal{L}_0^\dagger(\mathcal{O})\rangle_\ell+\langle\mathcal{L}^{\prime\dagger}(\mathcal{O})\rangle_{\ell-1},\;\;\ell\geq 1, 
\end{eqnarray}
where 
\begin{eqnarray}
\langle\mathcal{L}_0^\dagger(\mathcal{O})\rangle_\ell&=&\langle\text{i}\left[H_0,\mathcal{O}\right]+\mathcal{D}^\dagger(\mathcal{O})\rangle_\ell,\nonumber\\
\langle\mathcal{L}^{\prime\dagger}(\mathcal{O})\rangle_\ell&=&\langle \text{i}\left[H^\prime,\mathcal{O}\right]\rangle_\ell.
\end{eqnarray}
In this paper, we consider the external drive to be the perturbation, and carry out the analysis in the rotated frame (see Sec.~\ref{subsec:frame}). 

\subsection{Grading components of steady state}
\label{app:grading_steady_state}

In our calculations, we discard all expectation values of the many-body operators that are $\sim\mathcal{O}(|F|^\ell)$, with $\ell>2$. To determine the order of a specific many-body operator, we first grade different subspaces of the Hilbert space corresponding to the system, as follows. 

We work in the computational basis given by 
\begin{eqnarray}
    \ket{\mathbf{k}_n}=\left[\otimes_{i\in A}\ket{k_i}\right]\otimes\left[\otimes_{j\in B}\ket{k_j}\right]; \;k_{i},k_j\in\{0,1\},
\end{eqnarray}
where the total excitation number operator is defined as 
\begin{eqnarray}
    \mathcal{N}=\sum_{i\in A}a_i^+a_i^-+\sum_{j\in B}b_j^+b_j^-=\sum_n nP_n,
\end{eqnarray}
with the eigenvalue equation 
\begin{eqnarray}
    \mathcal{N}\ket{\mathbf{k}_n}=n\ket{\mathbf{k}_n}, n=0,1,2,\cdots
\end{eqnarray}
where 
\begin{eqnarray}
    n=\sum_{i\in A}k_i+\sum_{j\in B}k_j,
\end{eqnarray}
denotes the number of qubits in $\ket{\mathbf{k}_n}$ that are in the excited state $\ket{1}$. The projectors $P_n$ projecting a state on to a subspace spanned by $\ket{\mathbf{k}_n}$ corresponding to $n$ excitations are given by   
\begin{eqnarray}
    P_n=\sum_{\mathbf{k}_n}\ket{\mathbf{k}_n}\bra{\mathbf{k}_n}. 
\end{eqnarray}
We denote the state for which $n=0$ as $\ket{\mathbf{0}}$, referred to as the \emph{vacuum}.  The projectors 
satisfy $P_mP_n=0\;\forall\; m\neq n$,$P_n^\dagger=P_n^2=P_n$, and $\sum_nP_n=I$. In terms of these projectors, a density matrix $\tilde{\rho}$ in the rotated frame (see Sec.~\ref{subsec:frame}) can be written as
\begin{eqnarray}
    \tilde{\rho} &=&\sum_{m,n}\tilde{\rho}_{m,n},
    \label{eq:rho_decomposition_1}
\end{eqnarray}
with 
\begin{eqnarray}
    \tilde{\rho}_{m,n}=P_{m}\tilde{\rho} P_{n}
    \label{eq:rho_decomposition_2}
\end{eqnarray}
being the rectangular $(m,n)$th block of $\tilde{\rho}$ corresponding to rows belonging to the $m$ excitation sector, and columns belonging to the $n$ excitation sector. For $m\neq n$, the block $\tilde{\rho}_{m,n}$ collects the coherences between different excitation sectors. It is a null matrix if $\tilde{\rho}$ has no matrix element connecting the two sectors.

We decompose the Liouvillian 
\begin{eqnarray}
    \tilde{\mathcal{L}}(\tilde{\rho})=-\text{i}[\tilde{H}_S,\tilde{\rho}]+\mathcal{D}(\tilde{\rho})
\end{eqnarray}
as 
\begin{eqnarray}
    \tilde{\mathcal{L}}(\tilde{\rho})&=&\tilde{\mathcal{L}}_0(\tilde{\rho}) + \tilde{\mathcal{L}}_F(\tilde{\rho}),\nonumber\\
    &=&\left[\tilde{\mathcal{L}}_{00}(\tilde{\rho})+\tilde{\mathcal{L}}_{01}(\tilde{\rho})\right]+\tilde{\mathcal{L}}_F(\tilde{\rho}),
\end{eqnarray}
where
\begin{eqnarray}
    \tilde{\mathcal{L}}_{00}(\tilde{\rho})&=&-\text{i}[\tilde{H}_A+\tilde{H}_B+\tilde{H}_{AB},\tilde{\rho}],\nonumber\\
    \tilde{\mathcal{L}}_{01}(\tilde{\rho})&=&\mathcal{D}(\tilde{\rho}),\nonumber\\
    \tilde{\mathcal{L}}_F(\tilde{\rho})&=&-\text{i}[\tilde{H}_0,\tilde{\rho}].
\end{eqnarray}
Here, 
\begin{eqnarray}
    \mathcal{D}(\rho)=\sum_{i\in A}\gamma_i\mathcal{D}_i(\rho)+\sum_{j\in B}\gamma_{j}\mathcal{D}_j(\rho)+\Gamma \mathcal{D}_{AB}(\rho),
    \label{eq:cal_D_app}
\end{eqnarray}
with $\gamma_i,\gamma_j$, and $\Gamma$ ($\in \mathbb{R}$) being respectively the local and collective dissipation rates, and $\mathcal{D}_\alpha(.)$ are given by  
\begin{eqnarray}
    \mathcal{D}_\alpha(\rho)=\tilde{L}_\alpha \rho \tilde{L}_\alpha^\dagger-\frac{1}{2}\{\tilde{L}_\alpha^\dagger \tilde{L}_\alpha,\rho\},\;\alpha=i,j,AB,
\end{eqnarray} 
with $\tilde{L}_\alpha$ being the local ($\alpha=i,j$) and global $(\alpha=AB)$ jump operators. With the specific values of the parameters used in this paper, $\tilde{H}_{AB}=H_{AB}$, and the forms of $\tilde{L}_\alpha$ are given in Sec.~\ref{subsec:frame}.

Let us now define the \emph{grade} $s=m+n$ corresponding to a block $\tilde{\rho}_{m,n}$, and write $\tilde{\rho}_s\equiv\sum_{m+n=s}\tilde{\rho}_{m,n}$ for the grade-$s$ part of $\tilde\rho$ (keeping the labels $(m,n)$ whenever the individual block matters). Note that $s=0$ corresponds to both $m=0$ and $n=0$, designating all qubits in their ground states, i.e., the vacuum $\ket{\mathbf{0}}$. We now summarize the effects of different components of $\tilde{\mathcal{L}}(.)$ on a specific block $\tilde{\rho}_s$ of grade $s$ as follows. 
\begin{enumerate}
    \item Note that $[\mathcal{N},\tilde{H}_A+\tilde{H}_B+H_{AB}]=0$, implying the grade of $\tilde{\rho}_s$ remaining unchanged under the application of $\tilde{\mathcal{L}}_{00}$, i.e., $\tilde{\mathcal{L}}_{00}(\tilde{\rho}_s)\to\tilde{\rho}_s$. Note that $\tilde{L}_\alpha P_n=P_{n-1}\tilde L_\alpha$, and $[\mathcal{N},\tilde{L}_\alpha^\dagger\tilde{L}_\alpha]=0$. Further, since $\{\tilde{L}_\alpha^\dagger\tilde{L}_\alpha,\tilde{\rho}_s\}$ does not change the grade of $\tilde{\rho}_s$, and since
    \begin{eqnarray}
        \tilde{L}_\alpha\tilde{\rho}_s\tilde{L}^\dagger_\alpha&=&\tilde{L}_\alpha P_m\tilde{\rho}P_n\tilde{L}^\dagger_\alpha\nonumber\\
        &\to&P_{m-1}\tilde{\rho}P_{n-1}\equiv \tilde{\rho}_{s-2},
        \label{eq:grade_rules_1}
    \end{eqnarray}
    one obtains $\mathcal{D}(\tilde{\rho}_s)\to\tilde{\rho}_{s-2}$ (the anticommutation part preserves the grade, the jump part lowers it by two). Therefore, 
    \begin{eqnarray}
        \tilde{\mathcal{L}}_0(\tilde{\rho}_s)\to\tilde{\rho}_s,\tilde{\rho}_{s-2}.
        \label{eq:grade_rules_2}
    \end{eqnarray}
    \item On the other hand, 
    \begin{eqnarray}
        \tilde{\mathcal{L}}_F(\tilde{\rho}_{m,n})\to \tilde{\rho}_{m\pm 1,n},\mathrm{ or, }\tilde{\rho}_{m,n\pm 1}\equiv \tilde{\rho}_{s\pm 1}.
        \label{eq:grade_rules_3}
    \end{eqnarray}
\end{enumerate}

We now write the state $\tilde{\rho}$ as 
\begin{eqnarray}
    \tilde{\rho}=\sum_{\ell=0,1,2,\cdots}\tilde{\rho}^{(\ell)}, 
\end{eqnarray}
where $\tilde{\rho}^{(\ell)}$ collects all terms of exactly order $|F|^{\ell}$ in $\tilde{\rho}$, i.e., $\tilde{\rho}^{(\ell)}\sim\mathcal{O}(|F|^{\ell})$. For steady state 
\begin{eqnarray}
    \tilde{\rho}^\infty&=&\sum_{\ell=0,1,2,\cdots}\tilde{\rho}^{\infty,(\ell)}, 
    \label{eq:steady_state_expansion}
\end{eqnarray}
one can collect terms order-by-order in $|F|$ and write, from
$\tilde{\mathcal{L}}(\tilde{\rho}^\infty)=0$, 
\begin{eqnarray}
    \tilde{\mathcal{L}}_0(\tilde{\rho}^{\infty,(0)})&=&0,\nonumber\\
    \tilde{\mathcal{L}}_0(\tilde{\rho}^{\infty,(\ell)})+\tilde{\mathcal{L}}_F(\tilde{\rho}^{\infty,(\ell-1)})&=&0,\;\ell\geq 1.
    \label{eq:liouvillian_2}
\end{eqnarray}

Note, for the $0$th order contribution to the quantum master equation, given by 
\begin{eqnarray}
    \frac{d\tilde{\rho}^{(0)}}{dt}&=&\tilde{\mathcal{L}}_0\left(\tilde{\rho}^{(0)}\right),
\end{eqnarray}
that  $[\mathcal{N},\tilde{H}_A+\tilde{H}_B+\tilde{H}_{AB}]=0$. For a jump operator $\tilde{L}_\alpha$ that lowers $\mathcal{N}$ by one ($\mathcal{N}\tilde{L}_\alpha=\tilde{L}_\alpha(\mathcal{N}-1)$), the dissipator contributes
\begin{eqnarray}
    &&\mathrm{Tr}\left[\mathcal{N}\left(\tilde{L}_\alpha\tilde{\rho}^{(0)} \tilde{L}_\alpha^\dagger-\frac{1}{2}\{\tilde{L}_\alpha^\dagger \tilde{L}_\alpha,\tilde{\rho}^{(0)}\}\right)\right]\nonumber\\&=&\mathrm{Tr}\left[\left(\tilde{L}^\dagger \mathcal{N}\tilde{L}-\frac{1}{2}\{\mathcal{N},\tilde{L}^\dagger \tilde{L}\}\right)\tilde{\rho}^{(0)}\right]\nonumber\\&=&-\mathrm{Tr}[\tilde{L}_\alpha^\dagger \tilde{L}_\alpha\tilde{\rho}^{(0)}], 
\end{eqnarray}
since $\tilde{L}_\alpha^\dagger \mathcal{N}\tilde{L}=\tilde{L}_\alpha^\dagger \tilde{L}_\alpha(\mathcal{N}-1)$, and $\left[\tilde{L}_\alpha^\dagger \tilde{L}_\alpha,\mathcal{N}\right]=0$. Hence
\begin{eqnarray}
    \frac{d\langle \mathcal{N}\rangle^{(0)}}{dt}&=&\frac{d}{dt}\left[\mathrm{Tr}\left[\mathcal{N}\tilde{\rho}^{(0)}\right]\right]
    \nonumber\\&=&-\sum_{i\in A}\gamma_i\mathrm{Tr}\left[a_i^{+}a_i^-\tilde{\rho}^{(0)}\right]-\sum_j\gamma_j\mathrm{Tr}\left[ b_j^{+}b_j^-\tilde{\rho}^{(0)}\right]\nonumber\\&&-\Gamma\mathrm{Tr}\left[ \tilde{L}_{AB}^\dagger \tilde{L}_{AB}\tilde{\rho}^{(0)}\right]\nonumber\\&\le&-\gamma_{\min}\mathrm{Tr}\left[\mathcal{N}\tilde{\rho}^{(0)}\right],
    \label{eq:diff_decay}
\end{eqnarray}
where $\gamma_{\min}$ is the minimum of all local dissipation rates, and we have used $\mathrm{Tr}\left[ \tilde{L}_{AB}^\dagger \tilde{L}_{AB}\tilde{\rho}^{(0)}\right]\geq 0$. This leads to 
\begin{eqnarray}
    \langle \mathcal{N}\rangle^{(0)}(t)\leq \langle\mathcal{N}\rangle^{(0)}(0)\text{e}^{-\gamma_{\min}t}.
    \label{eq:decay}
\end{eqnarray}
The zeroth-order steady state $\tilde{\rho}^{\infty,(0)}$ is the $|F|\to0$ limit of the unique steady state, so it is positive with unit trace, and it satisfies $\tilde{\mathcal L}_0(\tilde{\rho}^{\infty,(0)})=0$. Further, while Eq.~(\ref{eq:decay}) suggests  $\lim_{t\to\infty}\langle\mathcal{N}\rangle^{(0)}(t)=\langle\mathcal{N}\rangle^{\infty,(0)}\leq 0$, $\langle\mathcal{N}\rangle^{\infty,(0)}$ must be $\geq 0$,  implying  $\langle\mathcal{N}\rangle^{\infty,(0)}=0$, and subsequently  $\tilde{\rho}^{\infty,(0)}=\ket{\mathbf0}\bra{\mathbf0}$  (note that this is the zeroth-order part only, the component $P_0\tilde{\rho}^{\infty}P_0$ of the \emph{full} steady state also contains the higher-order vacuum depletion). This corresponds to the vacuum $\ket{\mathbf{0}}$ as the unique zero mode of $\tilde{\mathcal{L}}_0$ (the \emph{only} grade $s=0$ state).   Therefore, Eq.~(\ref{eq:steady_state_expansion}) becomes 
\begin{eqnarray}
    \tilde{\rho}^{\infty}=\ket{\mathbf{0}}\bra{\mathbf{0}}+\tilde{\rho}^{\infty,(1)}+\tilde{\rho}^{\infty,(2)}+\cdots
\end{eqnarray}
where $\tilde{\rho}^{\infty,(\ell)}$ ($\ell\geq 1$) are traceless.

For ease of discussion, here onward, we include the grade in the subscript of the notation for components of steady state having different orders. For example, we write $\tilde{\rho}^{\infty,(0)}\equiv\tilde{\rho}^{\infty,(0)}_0$, and note that $\ket{\mathbf0}\bra{\mathbf0}$ is the only grade-$0$ operator, i.e.,  $\tilde{\rho}^{\infty,(0)}_0$ is the only grade-$0$ component at order $|F|^0$. At higher orders,  $\tilde{\rho}^{\infty,(\ell)}$ ($\ell=2,4,\dots$) also involves a grade-$0$ part $\propto\ket{\mathbf0}\bra{\mathbf0}$: it is the depletion of the vacuum population, fixed by $\mathrm{Tr}\,\tilde{\rho}^{\infty,(\ell)}=0$ (e.g.\ at order $|F|^2$ it equals $-\mathrm{Tr}\,\tilde{\rho}^{\infty,(2)}_{1,1}$). The second equation in (\ref{eq:liouvillian_2}) leads to a recursion relation 
\begin{eqnarray}
    \tilde{\rho}^{\infty,(\ell)}&=&-\tilde{\mathcal{L}}_{0}^{-1}\big(\tilde{\mathcal{L}}_F\big(\tilde{\rho}^{\infty,(\ell-1)}\big)\big).
    \label{eq:recursion}
\end{eqnarray}
Note that $\tilde{\mathcal{L}}_0$ has a \emph{unique} zero eigenvalue, with eigenvector $\ket{\mathbf0}\bra{\mathbf0}$. Note further that $\tilde{\mathcal{L}}_0$ is trace-preserving, mapping the space of traceless operators into itself. Since it excludes the zero mode,  $\tilde{\mathcal{L}}_0$ is invertible on traceless operators, and Eq.~(\ref{eq:recursion}) is justified, which determines the complete steady state order-by-order.

In the Liouvillian space organised by grades, $\tilde{\mathcal L}_0$ is block-triangular, with the grade-$s$ block mapping into grades $s$ and $s-2$ only (see Eq.~(\ref{eq:grade_rules_2})). For a traceless operator $X_s$ with grade $s$, 
\begin{equation}
    \tilde{\mathcal L}_0^{-1}(X_s)\to X_s,X_{s-2},X_{s-4},\dots
    \label{eq:inverse_grade}
\end{equation}
i.e., $\tilde{\mathcal L}_0^{-1}$, similar to $\tilde{\mathcal L}_0$, does not increase the grade. We now apply Eq.~(\ref{eq:recursion}) order-by-order, starting from $\tilde\rho^{\infty,(0)}\equiv\tilde\rho^{\infty,(0)}_0$, and note the following.
\begin{enumerate}
    \item $\tilde{\mathcal L}_F\big(\tilde\rho^{\infty,(0)}_0\big)$ has grade $1$ (Eq.~(\ref{eq:grade_rules_3})), and the grade stays $1$ due to Eq.~(\ref{eq:inverse_grade}).  Therefore,  $\tilde\rho^{\infty,(1)}\equiv\tilde\rho^{\infty,(1)}_1$.
    \item $\tilde{\mathcal L}_F\big(\tilde\rho^{\infty,(1)}_1\big)$ has grades $2$ and $0$ (see Eq.~(\ref{eq:grade_rules_3})). By Eq.~(\ref{eq:inverse_grade}), $\tilde\rho^{\infty,(2)}$ has maximum grade $2$, and the grade-$0$ part is $\propto\ket{\mathbf0}\bra{\mathbf0}$ with a coefficient fixed by $\mathrm{Tr}[\tilde\rho^{\infty,(2)}]=0$, i.e.,  $\tilde\rho^{\infty,(2)}=\tilde\rho^{\infty,(2)}_2+\tilde\rho^{\infty,(2)}_0$, with $\tilde\rho^{\infty,(2)}_0=-\mathrm{Tr}\big[\tilde\rho^{\infty,(2)}_{1,1}\big]\ket{\mathbf0}\bra{\mathbf0}$.
\end{enumerate}
More generally, if $\tilde\rho^{\infty,(\ell-1)}$ has the  maximum grade $\ell-1$,  (and grades $\ell-1,\ell-3,\dots$), then $\tilde{\mathcal L}_F\big(\tilde\rho^{\infty,(\ell-1)}\big)$ has maximum grade $\ell$ (and grades $\ell,\ell-2,\dots$), and by Eq.~(\ref{eq:inverse_grade}) so does $\tilde\rho^{\infty,(\ell)}$. Hence, by induction, we arrive at the following.
\begin{quote}
    For every $\ell\ge0$, $\tilde\rho^{\infty,(\ell)}$ has maximum grade $\ell$ and is supported only on the grades $\{\ell,\ell-2,\ell-4,\dots\}$ (up to $0$, or $1$). Equivalently, the blocks $\tilde\rho_{m,n}$ vanish at every order $|F|^\ell$ with $\ell<m+n$, so $\tilde\rho_{m,n}=\mathcal O(|F|^{m+n})$.
\end{quote}
Note that this provides an upper bound, i.e.,  a block of grade $s$ is at most $\mathcal O(|F|^s)$.

\subsection{Determining order of many-body operators}
\label{app:many_body_operator_order}

\begin{table}[t]
\caption{Minimal grades $m_{\min}$, $n_{\min}$ and $s_{\min}$ for products of one, two and three operators acting on different sites. An entry $\mathcal T_r$ ($\mathcal T_l$) means that the factor raises (lowers) $\mathcal N$ by one, e.g.,  $a_i^+$ or $b_j^+$ ($a_i^-$ or $b_j^-$), and empty cells indicate that the operator has fewer factors.}
\label{tab:minimal-grades}
\centering
\begin{tabular}{cccccc}
\hline\hline
$\mathcal T_1$ & $\mathcal T_2$ & $\mathcal T_3$ & $m_{\min}$ & $n_{\min}$ & $s_{\min}$ \\
\hline
\multicolumn{6}{c}{Single-term operator} \\
\hline
$\mathcal T_r$ &  &  & 0 & 1 & 1 \\
$\mathcal T_l$ &  &  & 1 & 0 & 1 \\
\hline
\multicolumn{6}{c}{Two-term operator} \\
\hline
$\mathcal T_r$ & $\mathcal T_r$ &  & 0 & 2 & 2 \\
$\mathcal T_r$ & $\mathcal T_l$ &  & 1 & 1 & 2 \\
$\mathcal T_l$ & $\mathcal T_r$ &  & 1 & 1 & 2 \\
$\mathcal T_l$ & $\mathcal T_l$ &  & 2 & 0 & 2 \\
\hline
\multicolumn{6}{c}{Three-term operator} \\
\hline
$\mathcal T_r$ & $\mathcal T_r$ & $\mathcal T_r$ & 0 & 3 & 3 \\
$\mathcal T_r$ & $\mathcal T_r$ & $\mathcal T_l$ & 1 & 2 & 3 \\
$\mathcal T_r$ & $\mathcal T_l$ & $\mathcal T_r$ & 1 & 2 & 3 \\
$\mathcal T_r$ & $\mathcal T_l$ & $\mathcal T_l$ & 2 & 1 & 3 \\
$\mathcal T_l$ & $\mathcal T_r$ & $\mathcal T_r$ & 1 & 2 & 3 \\
$\mathcal T_l$ & $\mathcal T_r$ & $\mathcal T_l$ & 2 & 1 & 3 \\
$\mathcal T_l$ & $\mathcal T_l$ & $\mathcal T_r$ & 2 & 1 & 3 \\
$\mathcal T_l$ & $\mathcal T_l$ & $\mathcal T_l$ & 3 & 0 & 3 \\
\hline\hline
\end{tabular}
\end{table}

We now consider expectation value of an operator $\mathcal{T}$ in $\tilde{\rho}^{\infty}$, given by (see Eqs.~(\ref{eq:rho_decomposition_1})-(\ref{eq:rho_decomposition_2}))
\begin{eqnarray}
    \langle\mathcal{T}\rangle^{\infty}=\mathrm{Tr}\left[\tilde{\rho}^{\infty}\mathcal{T}\right]=\sum_{n,m}\text{Tr}\left[P_m\tilde{\rho}^{\infty}P_n\mathcal{T}\right].
    \label{eq:exp_value}
\end{eqnarray}
By the cyclic property of trace, $\mathrm{Tr}\left[P_m\tilde{\rho}^\infty P_n\mathcal{T}\right]=\mathrm{Tr}\left[\tilde{\rho}^\infty P_n\mathcal{T}P_m\right]$, which vanishes unless $\mathcal{T}$ connects the $m$th and the $n$th sectors. For the remaining terms, $P_m\tilde{\rho}^{\infty}P_n=\tilde{\rho}_{m,n}\sim\mathcal{O}\left(|F|^{m+n}\right)$. Let $s_{\min}(\mathcal T)$ be the smallest grade $m+n$ among the pairs of sectors $(m,n)$ that $\mathcal T$ connects, i.e.\ with $P_n \mathcal T P_m\neq0$.   In $\langle\mathcal{T}\rangle$, the surviving terms are $\mathcal O(|F|^{m+n})$ with $m+n\ge s_{\min}(\mathcal T)$, and the smallest of these exponents is $s_{\min}(\mathcal T)$ itself. For small $|F|$, the term(s) with $m+n=s_{\min}(\mathcal T)$ dominate, leading to   
\begin{equation}
    \langle \mathcal T\rangle^\infty=\mathcal O\big(|F|^{s_{\min}(\mathcal T)}\big).
    \label{eq:smin}
\end{equation}
As an example, for $\mathcal T=a_i^+$ (raising $\mathcal N$ by one) only $n=m+1$ survives, so only the blocks $\tilde\rho^\infty_{0,1},\tilde\rho^\infty_{1,2},\dots$ appear in Eq.~(\ref{eq:exp_value}). Further, $\tilde\rho^\infty_{0,1}=\mathcal O(|F|)$, $\tilde\rho^\infty_{1,2}=\mathcal O(|F|^3)$, and so on, while $s_{\min}(a_i^+)=1$, such that $\langle a_i^+\rangle=\mathcal O(|F|)$. 

To extend this for products of operators, let 
\begin{eqnarray}
    \mathcal T=\mathcal T_1\mathcal T_2\cdots \mathcal T_L,
\end{eqnarray}
where each of the factors $\mathcal T_p\in\{a_i^+,a_i^-,b_j^+,b_j^-\}$, $p=1,2,\cdots,L$, and they act on different sites. It is sufficient to check whether $\mathcal{T}_p$ raises or lowers $\mathcal N$ by one, so in Table~\ref{tab:minimal-grades} we write $\mathcal T_r$ for a raising operator ($a_i^+$ or $b_j^+$) and $\mathcal T_l$ for a lowering operator ($a_i^-$ or $b_j^-$). Let $m_{\min}$ and $n_{\min}$ be the smallest $m$ and the corresponding $n$ with $P_n \mathcal T P_m\neq0$. A product with $r$ factors $\mathcal T_r$ and $l$ factors $\mathcal T_l$ ($r+l=p$) has to remove (add) $l$ ($r$) excitations, so it needs at least $l$ excitations to act on and ends with $n=m-l+r$. Hence
\begin{eqnarray}
    m_{\min}&=&l,\nonumber\\ n_{\min}&=&r,\nonumber\\ s_{\min}&=&m_{\min}+n_{\min}=p.
    \label{eq:smin_rule}
\end{eqnarray}
Table~\ref{tab:minimal-grades} lists $m_{\min}$, $n_{\min}$ and $s_{\min}$ for all choices of $\mathcal T_r,\mathcal T_l$ in products with $p=1,2,3$ factors, where each row follows from Eq.~(\ref{eq:smin_rule}). Therefore, a product of $p$ factors has $\langle \mathcal T\rangle=\mathcal O(|F|^p)$, and every operator with three or more factors can be ignored in a calculation up to order $|F|^2$.

Note that the information regarding the site on which each factor acts does not engage in the calculation as long as they are different. Factors acting on the \emph{same} site must be reduced first and written in normal order with at most one $a_i^+$ and one $a_i^-$ per site, followed by calculating $s_{\min}$ via counting the raising and lowering operators that remain, with the product $a_i^+a_i^-$ (and similarly $b_j^+b_j^-$) counted as two. For example, take $\mathcal T=a_i^-a_i^+b_j^+$, which, upon reduction, becomes $\mathcal T=(1-a_i^+a_i^-)b_j^+=b_j^+-a_i^+a_i^-b_j^+$, with a component $b_j^+$ having $s_{\min}=1$, and another component $a_i^+a_i^-b_j^+$ having $s_{\min}=3$. Therefore, in a calculation up to $\mathcal O(|F|^2)$, $\mathcal{T}$ is $\mathcal{O}(|F|)$. Note further that if $[\mathcal N,\mathcal T]=0$, only $m=n$ contributes, implying $m_{\min}=n_{\min}$ and $s_{\min}=2n_{\min}$, where $n_{\min}$ is the smallest $n$ with $\mathcal T P_n\neq0$. For example, $\langle a_i^+a_i^-\rangle$, $\langle b_j^+b_j^-\rangle$, as well as $\langle a_i^+a_{i'}^-\rangle$ and $\langle b_j^+b_{j'}^-\rangle$ have $n_{\min}=1$ and hence are are $\mathcal O(|F|^2)$, while $a_i^+a_i^-\,a_{i'}^+a_{i'}^-$ has $n_{\min}=2$, and hence is $\mathcal O(|F|^4)$.
 
\bibliography{references}

\end{document}